\RequirePackage{fix-cm}
\documentclass[twocolumn]{svjour3}          
\usepackage{nomencl}
\makenomenclature
\journalname{Astronomy Astrophysics Review}
\smartqed  
\usepackage{graphicx}
\usepackage{url}
\usepackage[caption=false]{subfig}
\usepackage{txfonts}
\usepackage[round]{natbib}

\newcommand{\apj}{Astrophys J }
\newcommand{\apjl}{Astrophys J Lett }
\newcommand{\aj}{Astronom J }
\newcommand{\apjs}{Astrophys J Suppl }
\newcommand{\mnras}{MNRAS }
\newcommand{\aap}{Astron Astrophys }
\newcommand{\aaps}{Astron Astrophys Suppl }
\newcommand{\aapr}{Astron Astrophys Rev }
\newcommand{\nat}{Nature }
\newcommand{\nar}{New Astronom Rev }
\newcommand{\pasa}{Publications Astron Soc Australia }
\newcommand{\pasp}{Publications Astron Soc Pacific }
\newcommand{\araa}{Annu Rev Astron Astr }
\newdimen\digitwidth
\setbox0=\hbox{2}
\digitwidth=\wd0
\catcode `#=\active
\newcommand#{\kern\digitwidth}
\begin{document}

\title{The faint radio sky: radio astronomy becomes mainstream
}


\author{Paolo Padovani}


\institute{P. Padovani \at
              European Southern Observatory, Karl-Schwarzschild-Str. 2,
D-85748 Garching bei M\"unchen, Germany \\
              Tel.: +49-89-32006478\\
              \email{ppadovan@eso.org}           
}

\date{Received: June 7, 2016 / Accepted: August 9, 2016}

\maketitle

\begin{abstract}
Radio astronomy has changed. For years it studied relatively rare sources, which emit
mostly non-thermal radiation across the entire electromagnetic spectrum,
i.e. radio quasars and radio galaxies. Now it is reaching such faint flux
densities that it detects mainly star-forming galaxies and the more common
radio-quiet active galactic nuclei. These sources make up the bulk of the
extragalactic sky, which has been studied for decades in the infrared,
optical, and X-ray bands. I follow the transformation of radio astronomy by
reviewing the main components of the radio sky at the bright and faint
ends, the issue of their proper classification, their number counts,
luminosity functions, and evolution. The overall ``big picture''
astrophysical implications of these results, and their relevance for a
number of hot topics in extragalactic astronomy, are also discussed. The
future prospects of the faint radio sky are very bright, as we will soon be
flooded with survey data. This review should be useful to all extragalactic
astronomers, irrespective of their favourite electromagnetic band(s), and
even stellar astronomers might find it somewhat gratifying.
\keywords{Radio continuum: galaxies \and Galaxies: active \and Galaxies:
  starburst \and Quasars: general \and Galaxies: statistics \and surveys}
\end{abstract}

\tableofcontents

\section{Introduction}\label{intro}

\begin{figure}
\centering
\includegraphics[width=8.4cm]{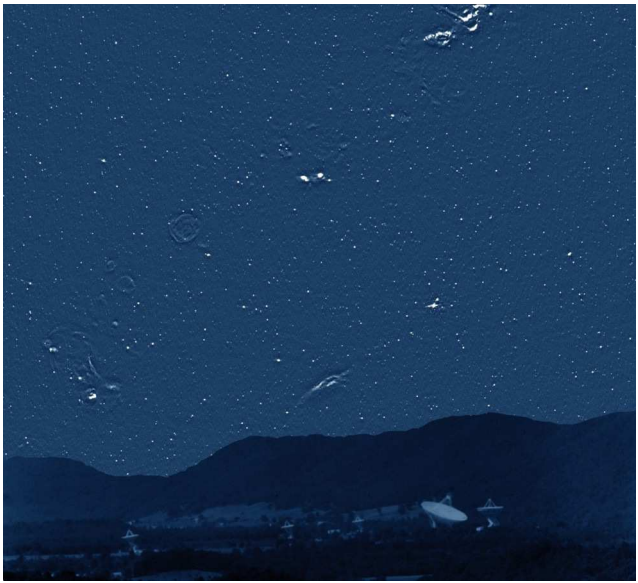}
\caption{The radio sky at 4.85 GHz above an optical photograph of the NRAO
  site in Green Bank, West Virginia (USA). The former 300-foot telescope
  made this image, which is about 45$^{\circ}$ across. Increasing radio
  brightness is indicated by lighter shades to indicate how the sky would
  appear to someone with a "radio eye" 300 feet ($\sim 91$ metres) in
  diameter. The flux density limit is $\sim 25$ mJy
  \citep{Gregory_1996}. Copyright NRAO.}
\label{fig:radionightsky}       
\end{figure}

The radio sky is very different from the optical sky. When we look at the
sky with the naked eye, we practically see only stars. These approximate
blackbody radiators and therefore their emission covers a relatively narrow
range of frequencies, centred at values ranging from the ultraviolet (UV)
to the infrared (IR), depending on the star temperature. Therefore, most
bright stars are extremely faint at radio frequencies. Figure
\ref{fig:radionightsky} displays the radio sky as it would appear to
someone with a ``radio eye'' with a diameter equal to that of the former
National Radio Astronomy Observatory (NRAO) Green Bank 300-foot radio
telescope. Most of the ``dots'' in the figure, which are unresolved radio
sources, are actually distant \cite[$z \approx 0.8$;][]{Condon_1989}
luminous radio galaxies (RGs) and quasars. The very few extended sources
are mostly Galactic supernova remnants. Radio emission from RGs and quasars
is due to ultra-relativistic ($E \gg m_{e} c^{2}$) electrons moving in a
magnetic field and thereby emitting synchrotron radiation, which, unlike
blackbody emission, can cover a very large range in frequency, reaching
$\sim 10$ decades in some sources.

\begin{figure}
  \centering
\subfloat[3C 31]{{\includegraphics[width=5.0cm]{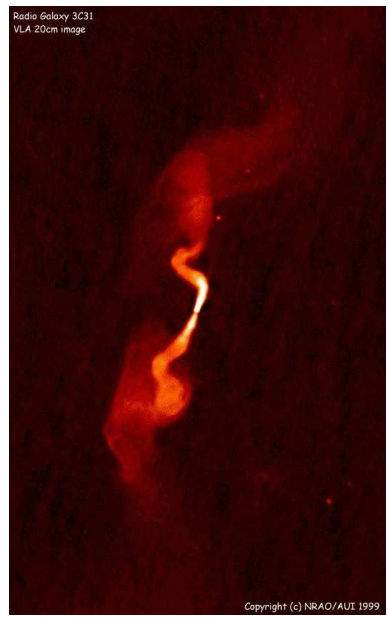} }}%
   \qquad
\subfloat[Fornax A]{{\includegraphics[width=5.0cm]{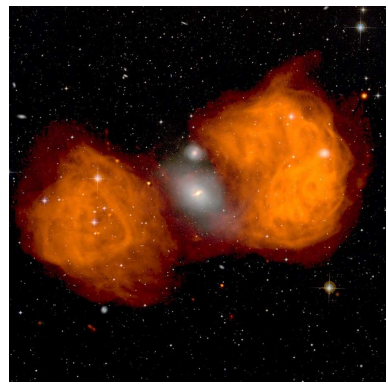} }}%
\caption{{\it (a):} large scale radio map at 1.4 GHz of 3C 31, showing
  filamentary plumes extending over 400 kpc from the galaxy. Copyright NRAO
  1996. Image from http://www.cv.nrao.edu/$\sim$abridle/images.htm. {\it
    (b):} Radio emission (orange) associated with the giant elliptical
  galaxy NGC1316 (centre of the image), consisting of two large radio
  lobes, each extending over $\sim 180$ kpc. Image courtesy of NRAO/AUI and
  J. M. Uson.}
\label{fig:3C_31_Fornax_A}
\end{figure}

As one goes fainter by using telescopes, galaxies take over even in
the optical sky: for AB $\gtrsim 20 - 22$ mag, depending on the filter,
galaxies outnumber stars by a large margin \citep{Windhorst_2011}. The
Hubble Ultra Deep Field (HUDF), which covers 11 arcmin$^2$ in four filters
($B$ to $z$) down to approximately uniform limiting magnitudes AB $\sim 29$
for point sources, contains at least 10,000 objects, almost all of them
galaxies \citep{Beckwith_2006}. Nevertheless, these galaxies are very
different from those seen in the radio ``bright''\footnote{The
  standard flux density unit in radio astronomy is the Jansky (Jy), which
  is equivalent to $10^{-23}$ erg cm$^{-2}$ s$^{-1}$ Hz$^{-1}$. By today's
  standards, strong radio sources have $S_{\rm r} \gtrsim 1$ Jy,
  intermediate ones have 1 mJy $\lesssim S_{\rm r} \lesssim 1$ Jy, while
  weak radio sources are below the mJy (soon $\mu$Jy) level.} sky. Radio
quasars and RGs, in fact, are somewhat rare, atypical, mostly non-thermal
sources across the entire electromagnetic spectrum, in which a large
fraction of the total emission comes from relativistic jets, that is
streams of plasma with speeds getting close to the speed of light, and 
associated lobes (Fig. \ref{fig:3C_31_Fornax_A}). Most of the galaxies detected in the
HUDF, on the other hand, are undergoing episodes of star formation
(SF) and therefore are strong thermal emitters.

This review\footnote{There was obviously no way I could mention {\it all}
  papers dealing with the many topics related to this review, which have
  appeared in the literature. I have therefore had to make choices and
  often resorted to the sentence ``and references therein''. Moreover, I
  here deal exclusively with extragalactic sources; see, e.g., Sect. 3.8 of
  \cite{Norris_2013} for a discussion of radio surveys of the Galactic
  plane.} aims to discuss very recent developments in our understanding of
the faint radio sky and how our radio view of the Universe has changed and
got much similar to the optical one. By going radio faint one is in fact
detecting the bulk of the active galactic nuclei (AGN) population, and not
only the small minority of radio quasars and RGs, and also plenty of
star-forming galaxies (SFGs). These developments are having (or should
have) a strong effect on radio astronomy, but should also change the
perception that astronomers working in other bands have of it. One of the
main messages of this review, in fact, is that radio astronomy is not
a ``niche'' activity but is extremely relevant also to more classical
aspects of extragalactic astronomy, such as star formation and galaxy
evolution.

I wrote this review primarily with non radio-astronomers in mind but I
believe that some radio astronomers might still be not fully aware of their
changing landscape and especially of what is in store for them. Therefore,
this review should be useful to all extragalactic astronomers irrespective 
of their preferred electromagnetic band(s). Stellar astronomers might still
find some satisfaction in learning that the faint radio sky is
dominated by SF related processes (which is another ``take
home'' message)!

The structure of this review is as follows: Sect. 2 discusses the main
astronomical components of the radio sky, while Sect. 3 deals with the
radio number counts. In Sect. 4 and 5 I describe the bright and faint radio
sky populations respectively, expanding in the latter on source
classification, radio number counts by class, luminosity functions (LFs),
and evolution. Sect. 6 puts these results into the bigger picture by
discussing some astrophysical implications, while Sect. 7 dwells on future
prospects by discussing upcoming radio facilities 
and their astrophysical impact, predictions
for deeper radio surveys, and the issue of source classification of very
faint ($\lesssim 1~\mu$Jy) radio sources. Finally, Sect. 8 gives my
conclusions and some messages. Throughout this paper, spectral indices are
defined by $S_{\nu} \propto \nu^{-\alpha}$, magnitudes are in the AB
system, and the values $H_0 = 70$ km s$^{-1}$ Mpc$^{-1}$, $\rm \Omega_{\rm
  m} = 0.27$, and $\rm \Omega_{\rm \Lambda} = 0.73$ have been used. The
state of the art of this field until early 2009 is reviewed by
\cite{deZotti_2010}.

\section{The main components of the radio sky}\label{sec:components}

I describe here the main properties of the sources, which populate the
radio sky,
namely galaxies (radio and star-form\-ing), radio quasars and blazars, and
radio-quiet AGN.  These properties are also 
extremely important for a proper classification of
faint radio sources (Sect. \ref{sec:class}). A recent
review of $z < 0.7$ AGN with radio powers\footnote{In this review I use W
  Hz$^{-1}$, i.e., power per unit frequency, which is commonly used in
  radio astronomy. It can be converted to erg s$^{-1}$ at 1.4 GHz ($\nu
  P_{\nu}$), for example, by multiplying by $1.4 \times 10^{16}$.} $P_{\rm
  1.4 GHz} > 10^{24}$ W Hz$^{-1}$ is given by \cite{Tadhunter_2016}.
  
\subsection{Radio galaxies}

Normal galaxies, with GHz radio powers 
$\lesssim 4 \times10^{20}$ W Hz$^{-1}$, are thought to have their radio
emission dominated by synchrotron radiation from interstellar relativistic
electrons \citep{Phillips_1986,Sadler_1989}. RGs are instead associated
with relativistic jets extending well beyond the host galaxy (see
Fig. \ref{fig:3C_31_Fornax_A}), which is typically a giant elliptical.
They are characterised by GHz radio powers $\gtrsim 10^{22}$ W Hz$^{-1}$
\citep[e.g.,][]{Sadler_1989,Ledlow_1996}, which represents the faint end of
the RG LF \citep[e.g.,][]{Urry_1995,Velzen_2012,Capetti_2015} and,
therefore, is a natural threshold for ``radio-loudness'' in galaxies
\citep[some papers have suggested the presence of non-thermal, jet emission
  in early type galaxies as faint as $10^{20}$ W Hz$^{-1}$:
  e.g.,][]{Balmaverde_2006}.  \cite{Fanaroff_1974} recognized that RGs
separate into two distinct luminosity classes, each with its own
characteristic radio morphology. High-luminosity Fanaroff-Riley (FR) IIs
have radio lobes with prominent hot spots and bright outer edges, while in
low-luminosity FR Is the radio emission is more diffuse
(Fig. \ref{fig:3C_31_Fornax_A}). The distinction is fairly sharp
at 178 MHz, with FR Is and FR IIs lying below and above, respectively, the
fiducial luminosity $P_{\rm 178 MHz} \approx 10^{26}/(H_0/70)^2$ W
Hz$^{-1}$. This translates to $P_{\rm 1.4 GHz} \approx 2 \times 10^{25}/(H_0/70)^2$
W Hz$^{-1}$, with some apparent dependency also on optical luminosity
\citep{Ledlow_1996}, which however appears not to be confirmed by more
recent studies \citep[][and references therein]{Gendre_2013}.

RGs are characterised by GHz radio spectra having $\alpha_{\rm r} \approx
0.7$. This is the signature of extended sources emitting synchrotron
radiation at relatively high frequencies where they are optically thin,
which implies the existence of fast electrons moving in a magnetic
field. Compact sources, instead, have flatter radio spectra, which are
attributed to synchrotron self-absorption \citep{Rybicki_2004}. More
specifically, different parts of the compact region become optically thick
at different frequencies, which results in a flattened integrated spectrum
over a relatively large range in frequency. A spectral index $\alpha_{\rm
  r} = 0.5$ divides in a remarkably clean way flat-spectrum/compact sources
from steep-spectrum/extended sources \citep[e.g.,][]{Massardi_2011}. An
exception to this rule is provided by compact steep-spectrum (CSS) radio
sources. These comprise $\approx 30\%$ of the bright radio source
population at a few GHz. Together with GHz peaked-spectrum (GPS), they are
generally considered to be young radio sources, which will eventually
evolve into large radio objects of the FR I and II type \citep[see][for 
  reviews]{O'Dea_1998,Sadler_2016}. Quite a few CSS/GPS sources are quasars.

The optical spectra of RGs are, as a rule, typical of so-called ``passive''
or ``quiescent'' galaxies, that is they display the absorption features
associated with an old stellar population, with some also revealing
powerful high ionization emission lines, and others showing at most weak
low-ionization emission lines. There appears to be in fact some other
fundamental differences between the two classes of RGs, such as their
emission line properties \citep{Hine_1979}, with FR IIs producing, for the
same radio power, $5 - 30$ times as much emission line luminosity
\citep{Zirbel_1995}. This has led to the suggestion that this dichotomy
might arise from differences in their central engines
\citep[e.g.,][]{Ghisellini_2001}, with jets produced by low accretion rate
sources being generally weak and mostly displaying FR I-type structure, and
galaxies with higher accretion rates giving rise to stronger, mainly FR
II-type jets. The environment appears also to have a role: radio sources in
rich clusters have a higher probability of being FR Is, which can be
explained by the fact that jets are more easily disrupted in dense
environments \citep[e.g.,][]{Gendre_2013}.

The different accretion rates have also been associated with the excitation
mode of the narrow line region gas in the host galaxy
\citep{Laing_1994}. In low-excitation RGs (LERGs) the accretion on to the
black hole is thought to be radiatively inefficient \citep{Chiaberge_1999},
while high-excitation RGs (HERGs), are instead linked to radiatively
efficient accretion discs of the type discussed by
\cite{Shakura_1973}. Following \cite{Heckman_2014} I will use the terms
``jet-mode'' and ``radiative-mode'' for the these two classes,
respectively. Note that radiative-mode radio-loud AGN include also (by definition) radio quasars
(see below). There is considerable overlap between jet-mode RGs and the
radio sources morphologically classified as FR Is and also between
radiative-mode RGs and FR IIs, although there is a sizeable population of
jet-mode FR IIs and a smaller one of radiative-mode FR I \cite[][and
  references therein]{Gendre_2013}. The two classes have also widely
different Eddington ratios\footnote{The ratio between the observed
  luminosity and the Eddington luminosity, $L_{\rm Edd} = 1.3 \times
  10^{46}~(M/10^8 \rm M_{\odot})$ erg/s, where $\rm M_{\odot}$ is one solar
  mass. This is the maximum luminosity a body can achieve when there is
  balance between radiation pressure (on the electrons) and gravitational
  force (on the protons).} with radiative-mode and jet-mode sources
typically above and below $L/L_{\rm Edd} \approx 0.01$ respectively 
\citep[e.g.,][]{Heckman_2014}.

\subsection{Radio quasars and blazars}\label{sed:quasars_blazars}

Radio quasars are intrinsically the same sources as some RGs. There is in
fact plenty of evidence indicating that they are simply FR
II/radiative-mode RGs with their jets at an angle $\lesssim 45^{\circ}$
w.r.t. the line of sight
\citep{Orr_1982,Barthel_1989,Antonucci_1993,Urry_1995}. The fact that radio
quasars display strong and Doppler broadened lines in their spectra (with
fullwidth half maximum [FWHM] $> 1,000$ km/s), unlike RGs, requires also
the presence of dust in a flattened configuration (the so-called
``to\-rus''), roughly perpendicular to the jet. Only when we look inside the
torus (and roughly down the jet) can we see the broad lines, emitted by
clouds moving fast close to the black hole, while for RGs the central
nucleus and surrounding material (including the broad line emitting clouds)
are obscured by the torus. The latter absorbs radiation along some lines of
sight re-emitting it in the IR. This so-called ``unification model''
explains in a natural way why the (projected) sizes of the jets of RGs are
larger than those of quasars \citep{Barthel_1989}. Radio quasars and FR
II/radiative-mode RGs can easily reach $P_{\rm 1.4 GHz} \approx 10^{27}$ W
Hz$^{-1}$ locally and $P_{\rm 1.4 GHz} \gtrsim 10^{28}$ W Hz$^{-1}$ at
higher redshifts \citep[e.g.,][]{Wall_2005,Padovani_2007}.

As regards FR I/jet-mode RGs, obscuration towards their nuclei appears to
be much smaller than that of their FR II/ra\-dia\-tive-mode relatives
\citep[e.g.,][]{Chiaberge_2002,Evans_2006}, which indicates that a torus
might be not present. This applies also to the population of FR II/jet-mode
RGs, which cannot be radio quasars seen at large angles
\citep{Hardcastle_2006}: their weak IR emission, in fact, suggests that,
like FR I/jet-mode sources, they also lack a torus
\citep{Ogle_2006}. Jet-mode RGs, therefore, irrespective of their radio
morphology, are ``unified'' with BL Lacertae objects (BL Lacs), a class of
AGN characterised by very weak, if any, emission lines. BL Lacs, together
with flat-spectrum radio quasars (FSRQs), make up the class of
blazars. FSRQs are defined by their radio spectral index at a few GHz
($\alpha_{\rm r} \le 0.5$), which, as mentioned above, is a sign of their
radio compactness. (Steep-spectrum radio quasars [SSRQs], not surprisingly,
have $\alpha_{\rm r} > 0.5$, extended radio emission, and jets that are at
angles w.r.t. the line of sight, which are intermediate between FSRQs and
FR II/radiative-mode RGs.) Blazars are AGN hosting jets oriented at a very
small angle ($\lesssim 15 - 20^{\circ}$) w.r.t. the line of sight. They
have very interesting and somewhat extreme properties, including
relativistic beaming, which leads to ``Doppler boosting'' of their flux
density (which makes blazars appear more powerful than they really are),
superluminal motion, large and rapid variability, and strong, non-thermal 
emission over the entire electromagnetic spectrum \citep{Urry_1995}
and possibly even beyond, into neutrino territory
\citep[e.g.,][]{Padovani_2015b,Padovani_2016}. The small angle their jets makes
w.r.t.  the line of sight implies that blazars are quite rare; nevertheless,
given their large flux densities, they are quite common in the bands
dominated by non-thermal sources (e.g., radio, sub-mm, and
$\gamma$-ray). For example, blazars constitute $\sim 51\%$ of the
classified sources in the 1 Jy 5 GHz catalogue \citep{Kuehr_1981}. Indeed,
the first quasar to be discovered, 3C 273 \citep{Schmidt_1963}, is an
FSRQ. And of the 2,023 $\gamma$-ray sources in the third {\it Fermi} Large
Area Telescope catalogue (3FGL) associated with an astronomical counterpart
$\sim 85\%$ of the total, and $\sim 98\%$ of extragalactic sources, are
blazars \citep{3FGL}. I refer the reader interested in the latest
developments on blazars and the subtleties of blazar classification to
\cite{Giommi_2012,Giommi_2013,Delia_2015} and references therein.

\subsection{Radio-quiet AGN}\label{sec:RQAGN}

Soon after the discovery of the first quasar, a very strong radio source
($S_{\rm 1.4GHz} \sim 50$ Jy), it was realised that there were many more
similar sources, which were undetected by the radio telescopes of the time:
they were ``radio-quiet'' \citep{Sandage_1965}.\footnote{Most of the
  so-called ``quasi-stellar galaxies'' described by \cite{Sandage_1965}
  actually turned out to be stars \citep[e.g.,][]{Kellermann_2015}; but the
  concept of radio-quiet quasars (i.e., the existence of quasars with much
  weaker radio emission) proved to be correct.} These sources were later
understood to be only ``radio-faint'', as for the same optical power their
radio powers were $\approx 3$ orders of magnitude smaller than their
radio-loud (RL) counterparts, but the name stuck. Radio-quiet (RQ) AGN,
which make up the majority ($> 90\%$) of the AGN class, were until
recently normally found in optically selected samples and are characterised
by relatively low radio-to-optical flux density ratios ($R \lesssim 10$)
and radio powers ($P_{\rm 1.4GHz} \lesssim 10^{24}$ W Hz$^{-1}$ locally:
Sect. \ref{sec:LF}).

Innumerable studies have compared the properties of the two AGN classes in
various bands to try to shed light on their inherent differences. As a
result, the distinction between the two types of AGN has turned out to be
not simply a matter of semantics. The two classes represent intrinsically
different objects, with RL AGN emitting a {\it large fraction} of their
energy non-thermally and in association with powerful relativistic jets,
while the multi-wavelength emission of RQ AGN is {\it dominated} by thermal
emission, directly or indirectly related to the accretion
disk\footnote{The words in italics highlight the presence of a thermal component
  (the UV bump, due to the accretion disk) in RL quasars and of a hot
  corona (producing the hard X-ray power law, due to inverse Compton of the
  optical/UV photons by high-energy electrons close to the disk) in RQ
  ones.}.
  
One of the strongest arguments in support of this statement comes from the
hard X-ray -- $\gamma$-ray bands. It is well established that, while many
RL sources emit all the way up to GeV ($2.4 \times 10^{23}$ Hz), and
sometimes TeV ($2.4 \times 10^{26}$ Hz), frequencies, RQ AGN have a sharp
cut-off at $E_{\rm c} \gtrsim 50$ KeV reaching $E_{\rm c} \approx 500$ keV
\citep[e.g.,][and references therein, where $F(E) \propto E^{-\Gamma}
  exp(-E/E_c)$]{Malizia_2014}. While $E_{\rm c}$ has been measured only in
a relatively small number of nearby bright Seyfert galaxies and its
determination is non-trivial (given the sensitivity required in the hard
X-ray band) such an exponential cut-off {\it must} be present in the
overall RQ AGN population at a few hundred keV in order not to violate the
X-ray background above this energy \citep[e.g.,][]{Comastri_2005}.
Furthermore, as I will discuss in Sect. \ref{sec:AGN_context}, no RQ AGN has so
far been detected in $\gamma$-rays; this means that, while RQ AGN are
actually only radio-weak, they are absolutely
$\gamma$-ray-quiet\footnote{At least based on current technology. One
  cannot exclude a scenario where, for example, the $\gamma$-ray flux in RQ
  AGN scales with $S_{\rm r}$ and therefore is $\approx 1,000$ times
  fainter than that of the RL sources detected by {\it Fermi}. Based on the
  relative numbers of RQ and RL AGN estimated in Table 1 of
  \cite{Padovani_2011a}, the RQ AGN contribution to the $\gamma$-ray
  background in this case {\it might} be non negligible, which would rule
  out this scenario, as there is no room left for other populations either
  than blazars, at least above 10 GeV \citep[e.g.,][]{Giommi_2015}.}. In short, high
energy observations do not allow the existence of a single class of AGN!
    
The host galaxies are also different. Those of RL AGN are bulge-dominated
($\rm L_{\rm bulge}/L_{\rm host} > 0.5$), i.e., ellipticals, while for RQ
ones the situation is more complicated. Luminous quasars ($M_{\rm B} \lesssim
-24$ or $L_{\rm bol} \gtrsim 10^{45}$ erg s$^{-1}$) at low redshifts are
mostly hosted by bulge-dominated galaxies
\citep[e.g.,][]{Dunlop_2003,Hopkins_2009a}, while lower luminosity sources
cover the full range of morphologies. However, quantitative knowledge of
the host properties of bright quasars at $z \gtrsim 0.5$ is still limited.
Finally, for many years it has been thought that the optical spectra of the
two classes were indistinguishable. This is not true, as long as spectra
with high enough signal-to-noise (S/N) ratio and resolution are available
\citep[e.g.,][and references therein]{Zamfir_2008,Sulentic_2011}. I discuss
this issue in some detail in Sect. \ref{sec:future_rl_rq}. 

As is the case for RL sources, unification applies also to RQ AGN, with
Type 2 AGN (e.g., Seyfert 2 galaxies), which show only narrow lines (with
FWHM typically $< 1,000$ km s$^{-1}$), having been unified with Type 1 AGN
(e.g., Seyfert 1 galaxies\footnote{It is generally understood that the
  distinction at $M_{\rm B} \sim -23$ used in the past to separate quasars
  and Seyferts, or stellar and non-stellar (i.e., extended) sources, is not
  a physical one. In this review I simply consider Seyfert 1's to be lower
  luminosity versions of quasars. Nevertheless, this absolute magnitude might
  still be useful to roughly separate AGN fainter and brighter
  than the brightest galaxies \citep[e.g.,][]{Condon_2013}.}), which
display broad lines. The former are then thought to be the same objects as
the latter with the central nucleus obscured by the
torus. \cite{Netzer_2015} discusses the latest developments for this
unified model, while \cite{Antonucci_2012} discusses both RL and RQ
unifications, with more emphasis on radio sources. RQ radiative-mode
sources are the ``classical'' broad- and narrow-lined AGN (Type 1 and 2),
while the jet-mode ones are the so-called LINERs \citep[see Table 4
  of][]{Heckman_2014}.

Radio emission in RQ AGN is relatively weak, unlike in RL AGN and RGs,
often spread across the host galaxy, and confined to the sub-kpc scale
\citep[e.g.,][]{Orienti_2015}. Most importantly, Seyferts and relatively
low redshift RQ quasars follow roughly the far-IR (FIR) -- radio
correlation (Sect. \ref{sec:SFGs}) typical of SFGs \citep[e.g.,][see also
  Sect. \ref{sec:emission_mech}]{Sopp_1991,
  Moric_2010,Sargent_2010}. Furthermore, the FIR flux density in Seyfert
galaxies correlates better with the low-resolution kpc-scale radio flux
density rather than with the high-resolution pc-scale emission
\citep{Thean_2001}, which points to a SF origin for the large scale radio
emission. This fits with the fact that low-luminosity RQ AGN are usually
hosted in late-type galaxies (Sect. \ref{sec:SFGs}). High resolution
studies using very long baseline interferometry (VLBI) imaging, on the
other hand, can reveal significant compact radio emission, often variable,
sometimes with evidence of weak jets, which can be linked to the central
AGN \cite[e.g.,][and references therein; these studies, until recently,
  could be carried out mostly for local sources;
  Sect. \ref{sec:emission_mech}] {Panessa_2013}.
  
\cite{Kimball_2011} have carried out sensitive ($S_{\rm 6GHz} \gtrsim 20~\mu$Jy) 
Very Large Array (VLA)
observations of 179 Sloan Digital Sky Survey (SDSS) quasars ($M_i < -23$)
with $0.2 < z < 0.3$ and found a peak in their luminosity distribution at
$P_{\rm 6 GHz} \approx 3 \times 10^{22}$ W Hz$^{-1}$. Their LF suggests
that low-redshift, low radio power quasars ($P_{\rm 6 GHz} < 10^{23}$ W Hz$^{-1}$
or $P_{\rm 1.4 GHz} \lesssim 4 \times 10^{23}$ W Hz$^{-1}$) are powered primarily
by SF and not by the central black hole.
These results have been confirmed by \cite{Condon_2013} using the NRAO VLA
Sky Survey (NVSS) and including also a high redshift ($1.8 < z < 2.5$)
sample.
 
\cite{Zakamska_2016} have recently come to the very different conclusion
that radio emission for the RQ quasars in their sample, which have $S_{\rm
  r} > 1$ mJy, $z < 0.8$, $4 \times 10^{21} \lesssim P_{\rm 1.4 GHz}
\lesssim 7 \times 10^{24}$ W Hz$^{-1}$, and $L_{\rm bol} > 10^{45}$ erg
s$^{-1}$, is dominated by quasar activity, not by the host galaxy. Their
result is based on a comparison between the observed radio power and the
value expected if radio emission were due to SF, using $L_{\rm FIR}$ to
derive the SF rate (SFR; see Sect. \ref{sec:astrop_SFG}). However, their $L_{\rm
  FIR}$ is derived from just one photometric point at $70~\mu$m or
$160~\mu$m using average calibrations, which give luminosities correct only
within an order of magnitude \citep{Symeonidis_2008}. Moreover, their
separation in RQ and RL AGN is done by applying a cut at $P_{\rm 1.4 GHz} = 7
\times 10^{24}$ W Hz$^{-1}$, which is on the high side for low-redshift RQ
AGN \citep[see, e.g.,][and discussions
  therein]{Padovani_1993,Padovani_2015a,Tadhunter_2016}.

It is fair to say that the mechanism responsible for the {\it bulk} of
radio emission in non-local RQ AGN has been a matter of debate for the past
fifty years or so, that is since their discovery. In addition, as mentioned
above, it is almost certain that more than one process is at play, with
contributions from both the central AGN and the host galaxy. I show later
on (Sect. \ref{sec:emission_mech}) that deep radio surveys play a big role
in sorting out this issue, which in my opinion has been basically solved
(at least at low powers).

In summary, the prevailing picture of the physical structure of AGN is
inherently axisymmetric and includes a black hole, an accretion disk,
which produces optical/UV and soft X-ray radiation, fast
moving clouds giving rise to the broad lines observed in AGN spectra,
gas, a torus, and slower moving clouds beyond the obscuring material
emitting narrower lines. Outflows of energetic particles can occur
along the poles of the disk or torus, escaping and forming collimated
jets and sometimes giant radio sources when the host galaxy
is an elliptical but forming only much weaker radio sources when the
host is a gas-rich spiral. This model implies a radically
different AGN appearance at different aspect angles. All of the above
applies to relatively large accretions rates: for $L/L_{\rm Edd}
\lesssim 0.01$ the disk becomes radiatively much less efficient, the
broad lines disappear, as does the evidence/need for a torus 
\citep[e.g.,][]{Heckman_2014}.

\subsection{Star-forming galaxies}\label{sec:SFGs}

SFGs can also be relatively strong radio emitters hosted by spiral and
irregular galaxies. These star-forming radio sources dominate the local ($z
< 0.4$) radio LF below $P_{\rm 1.4 GHz} \approx 10^{23}$ W
Hz$^{-1}$ but reach only $P_{\rm 1.4 GHz} \approx 10^{24}$ W Hz$^{-1}$
\citep[e.g.,][]{Sadler_2002,Mauch_2007}, as compared to $P_{\rm 1.4 GHz}
\approx 10^{27}$ W Hz$^{-1}$ for the powerful RGs. As is the case for RGs,
SFGs are also characterised by steep GHz radio spectra ($\alpha_{\rm r}
\approx 0.7$) dominated by synchrotron emission, but have also a flat free-free
component, which becomes predominant at $\nu \gtrsim 30$ GHz
\citep{Condon_1992}. Unlike RGs, however, where the ultimate prime mover is
the central black hole, in SFGs synchrotron emission results from
relativistic plasma accelerated in supernova remnants associated with
massive ($\rm M \gtrsim 8~M_{\odot}$) SF \citep{Condon_1992}. Radio
observations therefore probe very recent ($\lesssim 10^8$ yr) SF activity
and trace at some level its location as well. This is corroborated by one
of the tightest correlations in observational astrophysics, i.e., the FIR
-- radio correlation. FIR and radio emission are in fact strongly and
virtually linearly correlated in a variety of star-forming sources
\citep[e.g.,][and references therein]{Sargent_2010} and it is understood
that recent SF drives this correlation.

As regards their optical spectra, going from RGs to SFGs one moves along
the Hubble sequence from ellipticals to spirals and irregulars. This
implies several chan\-ges in the optical spectra, including a broad rise in
the blue continuum and a dramatic increase in the strengths of the nebular
emission lines, especially H$\alpha$ \citep{Kennicutt_1998}. All of the
above is however redshift dependent: the fraction of SFGs increases rapidly
with redshift \citep[e.g.,][]{Ilbert_2013} and indeed even RGs at $z
\gtrsim 3$ undergo vigorous SF \citep[e.g.,][]{Miley_2008}. In other words,
at high redshifts most galaxies are forming stars. For example, while SFRs 
of a few hundred $\rm M_{\odot}$ yr$^{-1}$ are very rare in
the local Universe and associated with so-called ``starburst (SB)
galaxies'', at high redshifts sustained SFRs are the norm, which means that
the concept of SB is a relative one. We know now, in fact, that the SF
activity of a galaxy can occur in two main different modes: a SB mode,
probably triggered by major mergers or dense SF regions; and a more normal
one, associated with secular processes, which is observed in the majority
of the SFGs \citep[][and references therein]{Bonzini_2015}. For this second
mode, a tight correlation between a galaxy's SFR and its stellar mass $\rm
M_{\star}$ has been discovered in the past decade
\citep{Noeske_2007,Elbaz_2007,Daddi_2007}. This is referred to as the
``main sequence'' (MS) of SFGs and is generally described as a single power
law of the form SFR $\propto \rm M_{\star}^{\beta}$ with $\beta \sim 0.4 -
1$ (depending on the sample). The slope of the correlation is approximately constant but the
normalization increases by about a factor of 20 from the local Universe to
$z \sim 2$ \citep[][and references therein]{Renzini_2015}.

A more modern definition of SFGs is therefore redshift dependent and
denotes galaxies, which belong to the MS, with SB and passive galaxies
being significantly above and below, respectively. That is, what is
relevant is not the absolute value of the SFR but its relative value
compared to the MS.

\subsection{FR 0 radio galaxies}\label{sec:FR0}

There is a relatively new entry in the radio sky: FR 0 RGs\footnote{This
  name was first used by \cite{Ghisellini_2010}.}! It turns out that $\sim
80\%$ of RGs in a SDSS/NVSS sample having $S_{\rm 1.4 GHz} > 5$ mJy and $z
\lesssim 0.3$ are unresolved or barely resolved at a radio resolution of 5
arcseconds \citep{Baldi_2010}: so no extended emission and therefore no FR
I/II division possible. The AGN power estimated from optical line or
core radio luminosities is at the same level of classical FR Is but the
sample shows a deficit of a factor $\sim 100$ in extended radio
power. Furthermore, $\sim 70\%$ of the local population of RGs at 20 GHz
studied by \cite{Sadler_2014} have also been classified as FR 0 since their
$\sim 1$ GHz emission is unresolved in the NVSS and Sydney University Molonglo 
Sky Survey (SUMSS) images.

Despite being called with the same name, I think there are some important
differences between the two samples. The 20 GHz sample, being selected at
high frequencies, is biased towards flat spectrum sources
(blazars and GPS; Sect. \ref{sec:obs_counts}); and indeed $\sim 61\%$ of
the FR 0s are candidate CSS/GPS sources, with $6 - 40\%$ of those being
possible blazars. Moreover, a good fraction of the radiative-mode FR 0s,
which in the 20 GHz sample are mostly hosted in late-type galaxies and make
up $\sim 25\%$ of the class, are very likely to be RQ AGN. This is because,
as discussed above, in these sources radio emission is very often confined
to the host galaxy, and therefore not very extended, but at the same time
there can be compact cores, which would be detected at high
frequencies. Indeed, 4/12 of the FR 0s studied by \cite{Baldi_2015} are RQ
AGN and Table 8 of \cite{Sadler_2014}, which lists the 13 spiral galaxies
with $z \le 0.025$ in their sample, includes many RQ AGN.  The remaining 20
GHz jet-mode FR 0s, which are not candidate CSS/GPS ($\sim 33\%$), are very
likely to be bona-fide RGs without the extended emission typically
associated with FR Is and IIs.

Some of the FR 0 ``RGs'' could also be low-redshift blaz\-ars misclassified
as RGs by current classification schemes because their non-thermal
radiation is not strong enough to dilute the host galaxy component even in
the Ca H\&K break region of the optical spectrum \citep[see][for more
  details]{Giommi_2012,Giommi_2013}. In other words, they would look like
LERGs in the optical band and like sources more core-dominated than classical FR
Is in the radio band, fitting the FR 0 definition. A study of the spectral
energy distributions (SEDs) of FR 0s from the radio to the high-energy
bands would easily pick out these sources, as their SEDs should be more
blazar- than RG-like.

To summarize, the study of FR 0 RGs is still in its infancy but 
it looks like this class could be quite heterogeneous. If it
will be confirmed that the large majority of RGs are true FR 0s, i.e., they
look like jet-mode FR Is without the extended radio
emission, we will need to figure out why this is the case and what this
means, for example, for unified schemes of RL sources.

\section{Radio number counts}

\subsection{Number count basics}\label{sec:counts_basics}

The simplest thing one can do when studying a flux-limited sample of
astronomical sources is to count them. This requires no additional
data but nevertheless number counts provide very useful information, as
their shape is tightly related to the evolutionary properties of the
sources and also to the geometry of the Universe (since the volume is not
simply $\propto$ to D$^3$ but also depends on the curvature). This is
illustrated by the following equation, which gives the differential number
counts (i.e., the number of objects per flux density per steradian):

$$\frac{n(S)}{4 \pi} = $$
\begin{equation}
4 \pi \frac{c}{H_0} \int_{z_{\rm min}(S)}^{z_{\rm
    max}(S)} \frac{\Phi[P(S,z),z] D_{\rm L}^4(z) dz}
     {(1+z)^{(3-\alpha)} \sqrt{(1+z)^2 (1+ \Omega_{\rm m}z) -
         z(z+2)\Omega_{\rm \Lambda}}},
\label{eq:counts}
\end{equation}

where $c$ is the speed of light, $\Phi(P,z)$ is the redshift dependent LF
(number of sources per unit power per unit comoving volume), $D_{\rm L}(z)$ is the
luminosity distance, and $z_{\rm min}(S)$ and $z_{\rm max}(S)$ represent
the flux density dependent redshift range over which the integration is
carried out. Eq. \ref{eq:counts} shows how the cosmological model (through
$H_0$, $\Omega_{\rm m}$, $\Omega_{\rm \Lambda}$, and $D_{\rm L}(z)$) and
the shape and evolution of the LF play a role in building the number
counts. One can parametrize the LF evolution in a simple way by writing
\citep[e.g.,][]{deZotti_2010}\footnote{This equation differs from eq. 11 in
  \cite{deZotti_2010} by a factor $1/f_{\rm L}(z)$, since they define their
  LFs as $\Phi(\rm log P$); see Sect. \ref{sec:LF}.}:

\begin{equation}
\Phi(P,z) = \Phi(P/f_{\rm L}(z),z=0) f_{\rm D}(z) /f_{\rm L}(z),
\label{eq:evol}
\end{equation}

which allows for changes in both power and number. The two extreme cases
are: 1. $f_{\rm D}(z) = 1$, which means that the comoving number density is
constant and $P(z) = P_0 f_{\rm L}(z)$, the so-called pure luminosity
evolution (PLE) case; 2. $f_{\rm L}(z) = 1$, which implies a constant power
and a density evolution $\Phi(z) = \Phi_0 f_{\rm D}(z)$, the so-called pure
density evolution (PDE) case. In more complex cases both $f_{\rm D}$ and
$f_{\rm L}$ can also have a dependence on power.  One normally talks about
``positive'' or ``negative'' evolution if $f(z) > 1$ or $< 1$ respectively,
meaning that, in the first case, for example, the power or the number
density was larger at higher redshifts\footnote{In terms of cosmic time,
  rather than redshift, it should be the other way around. Nevertheless, this is how
  these terms are generally used.}.

\begin{figure}
\centering
\includegraphics[width=8.4cm]{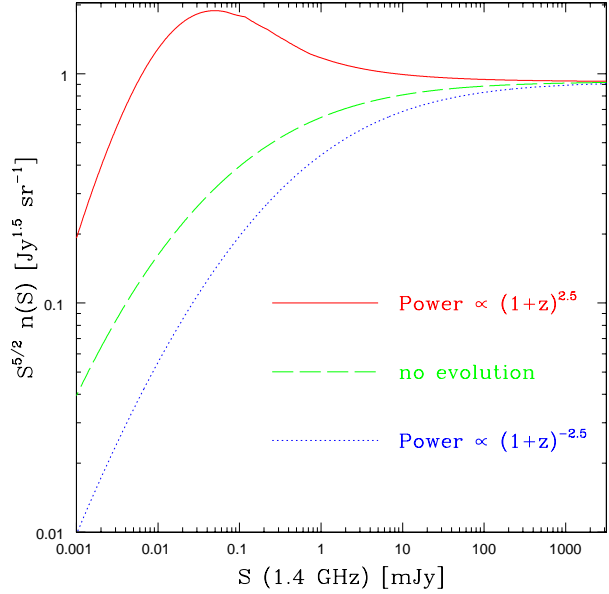}
\caption{Euclidean normalized 1.4 GHz source counts derived for different
  luminosity evolutions up to $z=2$ (and constant thereafter): $\propto
  (1+z)^{2.5}$ (solid red line), no evolution (dashed green line), and
  $\propto (1+z)^{-2.5}$ (dotted blue line). The local LF is that of
  \cite{Mauch_2007}; $\alpha_{\rm r} = 0.7$ was assumed.}
\label{fig:counts_exa}       
\end{figure}

Radio astronomers plot their differential number counts normalized by the
counts expected in a static Euclidean Universe\footnote{These can be simply
  derived under the assumption of a uniform distribution of sources and
  Euclidean space since the number of sources is $\propto D^3 \equiv
  (L/4\pi S)^{3/2}$. The integral counts, that is the number of objects
  seen on the sky with flux density $>S$ will then be $n(\ge S) \propto
  S^{-3/2}$, which translates into differential counts $n(S) \propto
  S^{-5/2}$.}, $n_E(S) \propto S^{-5/2}$, by displaying the quantity
$S^{5/2} n(S)$.
  
Figure \ref{fig:counts_exa} shows an example of Euclidean normalized 1.4 GHz
source counts under different evolutionary assumptions. For simplicity (and
based on real data: see Sect. \ref{sec:evolution}) I have taken the case of
a PLE, with $P(z) = P_0 (1+z)^{k}$, up to $z=2$ (and constant thereafter).
I have considered three cases: $P(z) \propto (1+z)^{2.5}$ (positive
evolution [$k > 0$]; solid red line), no evolution ($k=0$; dashed green
line), and $\propto (1+z)^{-2.5}$ (negative evolution [$k < 0$]; dotted
blue line). The local LF for SFGs of \cite{Mauch_2007} has been adopted,
and $\alpha_{\rm r} = 0.7$ was assumed.

Figure \ref{fig:counts_exa} shows the following: 1. the counts are Euclidean
and evolution independent only at large flux densities ($S \gtrsim 0.1$ Jy
in this example); 2. at fainter flux densities the geometry of the Universe
starts to have an effect, as clearly visible for the no-evolution case,
which soon diverges from the Euclidean one; 3. strong positive evolution
manages to counterbalance this, at least down to $\sim 0.1$
mJy\footnote{This value depends on the LF, the assumed evolution, and on
  the redshift at which evolution stops.}, while negative evolution makes
this effect more prominent. In short, once the cosmological model and
spectral index are fixed, the number counts are strongly dependent on
source evolution, but not only: the same counts, in fact, can be obtained
by different combinations of (local) LF and evolution.

\subsection{Observed radio number counts and their complications}\label{sec:obs_counts}

\begin{figure}
\centering \includegraphics[width=8.4cm]{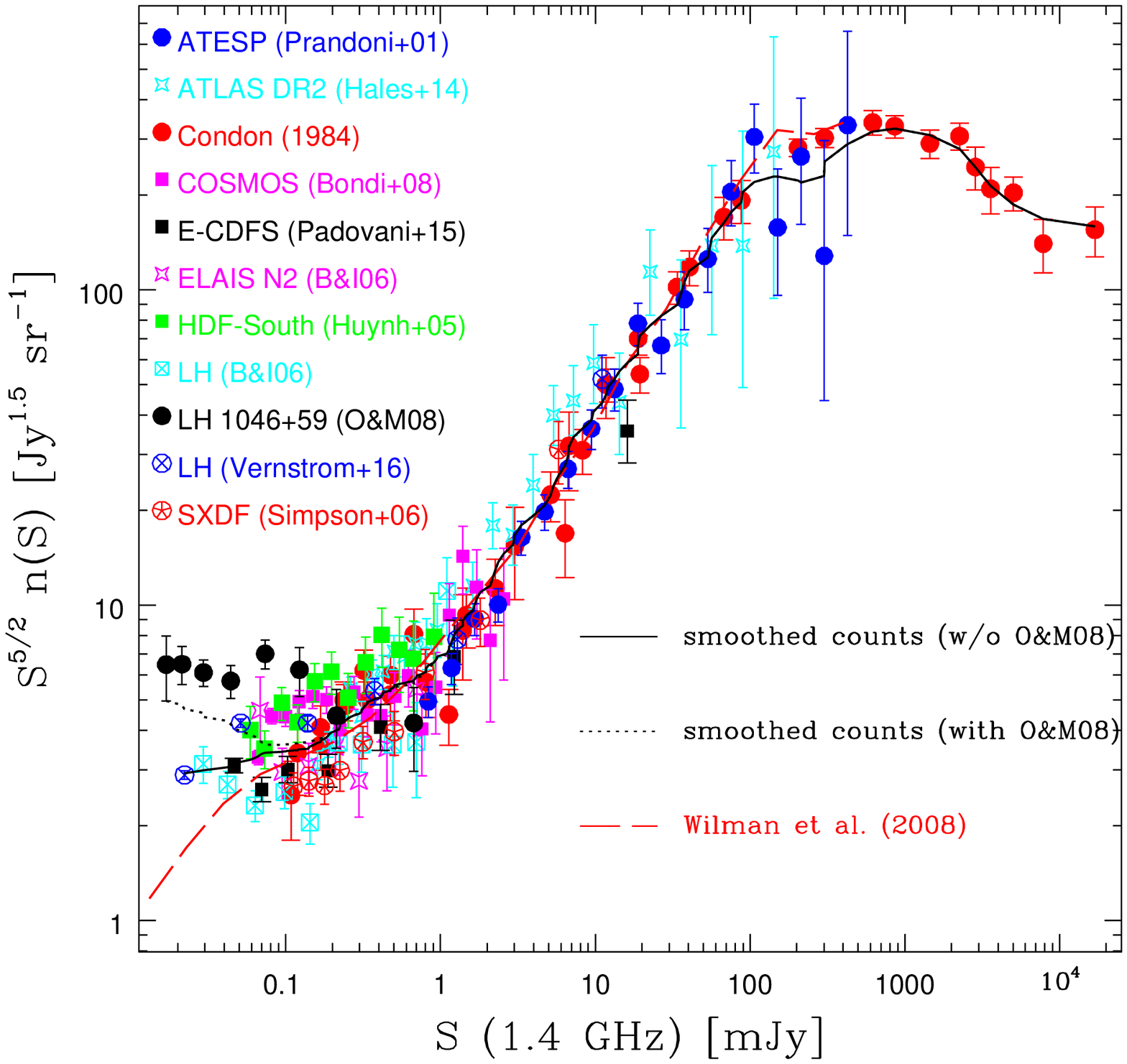}
\caption{The Euclidean normalized 1.4 GHz source counts derived from
  various surveys, as indicated in the legend: \cite{Prandoni_2001},
  \cite{Hales_2014b}, \cite{Condon_1984a}, \cite{Bondi_2008},
  \cite{Padovani_2015a}, \cite{Biggs_2006} (LH stands for Lockman Hole),
  \cite{Huynh_2005}, \cite{Owen_2008}, \cite{Vernstrom_2016} (converted
  from 3 GHz assuming $\alpha_{\rm r} = 0.7$), \cite{Simpson_2006}. The
  solid and dotted black lines are a smoothed version of the counts
  respectively excluding and including the \cite{Owen_2008} sample. The red
  dashed line represents the simulated number counts from the SKA Design
  Study \citep[SKADS;][]{Wilman_2008}. See \cite{Smolcic_2015} for even
  more survey data points. Most data and the SKADS counts courtesy of Isabella
  Prandoni.}
\label{fig:counts_obs}       
\end{figure}

Figure \ref{fig:counts_obs} shows the observed Euclidean normalized 1.4 GHz
source counts based on a variety of surveys, which reach $\sim
15~\mu$Jy. Based on Fig. \ref{fig:counts_exa}, we can immediately infer the
following: 1. strong radio sources have a pronounced positive evolution;
2. their LF is shifted to much higher powers ($\approx$ three orders of
magnitude assuming similar evolutions) as compared to the SFG LF used in
Fig. \ref{fig:counts_exa}, since the counts peak at a much larger flux
density; 3. below $\sim 1$ mJy a new population is very likely to make its
appearance, as the counts show a marked flattening\footnote{This is true
  only for the Euclidean normalized counts: the differential counts
  actually steepen.}, which one does not expect for normal, well-behaved
LFs; 4. below $\sim 1$ mJy, however, the observed source counts appear also
less well constrained with a scatter between different surveys, which can
reach a factor of $\approx 3$. I defer discussions of points $1 - 3$ to
Sect.  \ref{sec:deep_pop} and elaborate here somewhat on point
4\footnote{The subject of the ``proper'' estimation of radio number counts
  is a very complicated one, fraught with many issues, which go beyond the
  main scope (and length) of this review \citep[see, for
    example,][]{Condon_2012,Hales_2014,Padovani_2015a}. Here I briefly
  touch upon it.}.

Undoubtedly, some fraction of the scatter at faint radio flux densities may
be due to cosmic variance\footnote{This is the uncertainty in observational
estimates of extragalactic objects arising from the underlying large-scale
density fluctuations, which is often significant, especially in deep
surveys, which tend to cover relatively small areas.}. Nevertheless, this cannot
be the full story: for example, in the X-ray band, the scatter between the
(integral) number counts in the Great Observatories Origins Deep Survey
(GOODS) North and South fields, which cover $\sim 0.1$ deg$^2$, is only
$\sim 25\%$ at the faintest fluxes ($\approx 3\sigma$), fully consistent
with small field-to-field variations \citep{Luo_2008}. Published deep radio
source counts, on the other hand, are widely discrepant, even when made
with the same instrument, and even with different researchers using the
same instrument in the same field \citep{Condon_2012}! The central problem
in understanding radio source counts is the proper balance between source
confusion at lower resolution (many faint sources crowding into a single
beam) and missing lower surface brightness sources or underestimating 
flux densities at high resolution (i.e., resolving them out).
These issues, while better understood than in the 1950s (see below), still
exist today. There is no ``correct'' resolution that avoids confusion and
resolving out sources, but there is an optimum resolution depending on the
areal density of sources. As discussed by \cite{Norris_2013}, some of the
largest differences between the various surveys can also be accounted for
by different ways of handling many necessary corrections for effects such
as, to name a few: clean bias, Eddington bias, and completeness
corrections, together with imaging errors such as excessive deconvolution,
bandwidth smearing, and insufficient beam sampling in the image
plane. Furthermore, some authors measure the number of radio components,
while others measure the number of sources, each of which may consist of
several components, the numbers of which in turn may vary as a function of
flux density. Resolving these discrepancies is quite critical (but
difficult), since this scatter introduces uncertainties in the comparison
of observed number counts with detailed, model-based predictions.

I have concentrated on number counts around a few GHz because this is
where at present we can reach fainter flux densities. Angular resolution
scales with $\lambda$, therefore low-frequencies are penalized in this
respect. The beam solid angle of a radio telescope scales as $\nu^{-2}$ and
system noise generally increases with frequency, so the time needed to
survey a fixed area of sky to a given limit rises very rapidly at higher
frequencies. For example, the deepest image at 150 MHz 
recently obtained with the LOw Frequency ARray (LOFAR) goes down to $\sim 0.7$ mJy
\citep{Williams_2016} \citep[see also][for a deep 62 MHz field]{vanWeeren_2014}, 
equivalent to $S_{\rm 1.4GHz} \sim 0.15$ mJy (assuming $\alpha_{\rm r} = 0.7$), 
while the deepest 15.7 GHz survey reaches 0.1 mJy \citep{Whittam_2016},
which corresponds to $S_{\rm 1.4GHz} \sim 0.1 - 0.5$ mJy (for $\alpha_{\rm
  r} = 0 - 0.7$, where the first value is more appropriate for flat
spectrum cores). These values are $\gtrsim 10$ times larger than the
faintest 1.4 GHz flux density limit. It would be good to reach depths
comparable to the $\sim$ GHz surveys at lower and higher frequencies to
further constrain the number counts and get different and complementary
views on radio sources. Steep synchrotron spectra objects, i.e., SFGs and
RGs, are better detected at low frequencies, since their flux densities
increase rapidly $\propto \nu^{-\alpha_{\rm r}}$. High frequency surveys,
on the other hand, will be by default more biased towards flat spectrum
sources i.e., blazars and quasars in general
\cite[e.g.,][]{Giommi_2009,Mahony_2011,Whittam_2015}.  This also explains
why, despite reaching similar equivalent 1.4 GHz flux densities, the 150
MHz normalized counts display a flattening below 10 mJy ($S_{\rm 1.4GHz}
\sim 2$ mJy) while the 15.7 GHz ones do not (see Sect. \ref{sec:deep_pop}).

In closing this section, I would like to stress the relevance radio number
counts have had for cosmology and the Steady-State vs. Big Bang debate. It
was in fact the steepness of the earliest radio source counts, which gave
the first indications of cosmic evolution \citep[][see
  Fig. \ref{fig:counts_exa}]{Ryle_1955} but these same results led to the
Sidney-Cambridge controversy\footnote{The interested reader can find a
  detailed account of this controversy in \cite{Sullivan_1984}.}  in the
1950s over the nature of radio sources and their role in cosmology
\citep{Mills_1958}, which revolved around understanding source
confusion. Note that this was well before the discovery of quasars in 1963!
Sadly, as stated by McCrea in \cite{Sullivan_1984} ``In retrospect, in
spite of the confusing side issues, from about 1955 cosmologists would have
been safe in accepting that radio astronomy had shown the actual Universe
to be not in a steady state. Instead, they waited until a decade later when
the discovery of the microwave background had confirmed a positive
prediction of big-bang cosmology''.

\section{The bright radio sky population}\label{sec:bright_pop}

The study of the radio sky goes all the way back to the end of the
1940s, when \cite{Bolton_1949} identified three of the strongest radio
sources in the sky, namely Taurus A\footnote{In those early days radio
  sources were named after the constellation in which they appeared
  followed by a letter. Thus, Taurus A was the first object discovered in
  the Taurus constellation.}, Virgo A, and Centaurus A. Bolton et
al. associated Taurus A with the Crab Nebula, already known to be the
expanding shell of SN 1054. They also correctly identified the other two
sources with M 87 and NGC 5128 (both FR Is) but, realizing that, 
were they extragalactic their radio power would be enormous (for the time), they
concluded that if the identifications were correct it would imply that M 87
and NGC 5128 had to be within our own Galaxy. Bolton later explained that
he understood the true nature of the two sources ``but that he was
concerned that a conservative Nature referee might hold up publication''
\citep{Kellermann_2015}! By the mid 1950s, however, many RGs
were identified with optical counterparts and most high Galactic latitude
sources were recognized to be extragalactic with radio powers $\approx 10^7
- 10^{10}$ times larger than that of the Crab Nebula \citep[e.g.,
  $8 \times 10^{42}$ erg s$^{-1}$, i.e., $\approx 10^{27}$ W Hz$^{-1}$ at
  100 MHz in the case of Cygnus A:][]{Baade_1954}.

The bright radio sky turned out to be made up almost exclusively of RGs and
radio quasars. For example, the second revision of the Third Cambridge
Catalogue of Radio Sources (3CRR) \citep{Laing_1983}, which includes all
objects with $S_{\rm 178MHz} \ge 10$ Jy, $\delta \ge + 10^{\circ}$, and
$|b_{\rm II}| \ge 10^{\circ}$, contains only RGs and RL quasars. And out
of the 527 1 Jy 5 GHz sources \citep{Kuehr_1981} only two, NGC 1068 (a
Seyfert 2) and M 82 (a starburst), both at very low redshift ($z \le 0.004$), do
not belong to the RG, radio quasar, or blazar classes.

\begin{figure}
\centering
\includegraphics[width=8.4cm]{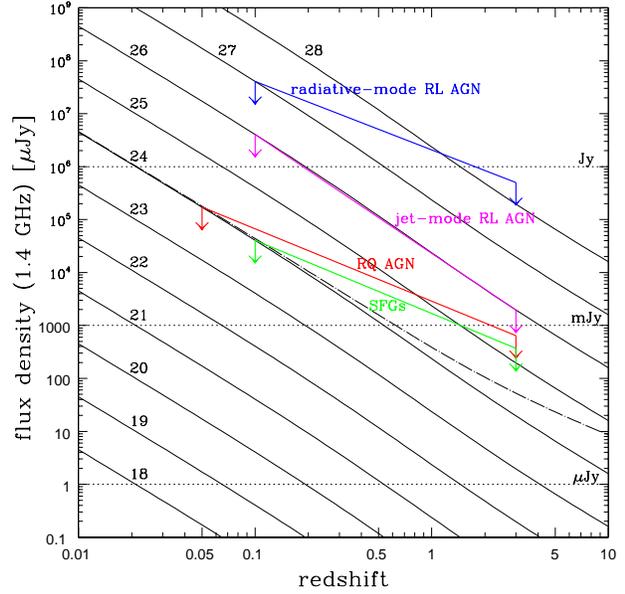}
\caption{Flux density at 1.4 GHz vs. redshift for various radio powers (in
  units of log W Hz$^{-1}$) assuming $\alpha_{\rm r} = 0.7$ (solid lines). For
  $P_{\rm 1.4GHz} = 10^{24}$ W Hz$^{-1}$ the case $\alpha_{\rm r} = 0$ is
  also shown (dot-dashed line). The flux density values corresponding to
  the maximum radio powers for various classes at $z \sim 0.05 - 0.1$ and
  $z \sim 3$ are also shown, as detailed in the text. The dotted
  horizontal lines indicate the 1 $\mu$Jy, 1 mJy and 1 Jy flux densities.}
\label{fig:flux_z}       
\end{figure}

The reason why this is the case is shown graphically in
Fig. \ref{fig:flux_z}, which plots the 1.4 GHz flux density vs. redshift
for a range of radio powers assuming $\alpha_{\rm r} = 0.7$ (solid lines;
for $P_{\rm 1.4GHz} = 10^{24}$ W Hz$^{-1}$ the case $\alpha_{\rm r} = 0$ is
also illustrated). It also shows the flux density values corresponding to
the maximum radio powers for radiative-mode and jet-mode RL AGN (blue and
magenta lines), RQ AGN (red line), and SFGs (green line) at $z \sim 0.05 -
0.1$ and $z \sim 3$\footnote{The local maximum powers are from
  \cite{Best_2014} (radiative- and jet-mode RL AGN), \cite{Padovani_2015a}
  (RQ AGN), and \cite{Mauch_2007} (SFGs). The values at $z \sim 3$ have
  been estimated by using luminosity evolution models from
  \cite{Urry_1995,Padovani_2015a}, and Padovani et al., in preparation}. 
  Figure \ref{fig:flux_z} gives a
simple, order of magnitude, picture of the largest flux densities reached
by the classes of sources, which populate the radio sky. The average
luminosities, being generally close to the break in the LF, and the related
more typical flux densities, are much smaller than the maximum ones.
 
Figure \ref{fig:flux_z} show the obvious dimming due to the dependence on the
inverse square of the luminosity distance, which can be partly offset by
luminosity evolution (and some k-corrections). The most important, less
trivial, message, is that different classes will cover widely different
ranges of flux densities, with RL AGN being essentially the only
``inhabitants'' of the (GHz) $\gtrsim 10 - 100$ mJy sky, apart from
SFGs and RQ AGN at very low redshift ($z \lesssim 0.02$), as indeed
observed. I now move to the focus of this review, that is the faint radio sky. 

\section{The faint radio sky population}\label{sec:deep_pop}

A turning point in the study of the radio sky was the realisation around
1984 that the Euclidean normalized 1.4 and 5 GHz source counts exhibited a
significant flattening below $\approx 1$ mJy
\citep{Condon_1984,Fomalont_1984, Windhorst_1984}. Furthermore,
\cite{Windhorst_1985}, based on optical identifications available for less
than half of the sample, suggested that ``for $1 < S_{\rm 1.4} < 10$ mJy a
blue radio galaxy population becomes increasingly important; these often
have peculiar optical morphology indicative of interacting or merging
galaxies''. To really understand which sources were responsible for the
flattening and sort out the source population of the $\lesssim 1$ mJy radio
sky took more than thirty years. I discuss why next.

\subsection{Source classification}\label{sec:class}

A problem common to all astronomical surveys is that of the classification
of sources. After having detected them, in fact, one wants to figure out
what they are, which is vital to extract astrophysical information. This
requires a determination of the redshift, without which the emitted power
cannot be calculated, which is still mostly done through optical/near-IR
(NIR) spectroscopy\footnote{The Atacama Large Millimeter/submillimeter
  Array (ALMA) can also determine redshifts of, for example, SFGs through
  their molecular emission lines at millimetre wavelengths
  \citep[e.g.,][]{Weiss_2013}.}.

\begin{figure}
\centering
   \vspace{-2.5em}
\subfloat[SFG]{{\includegraphics[width=8.0cm]{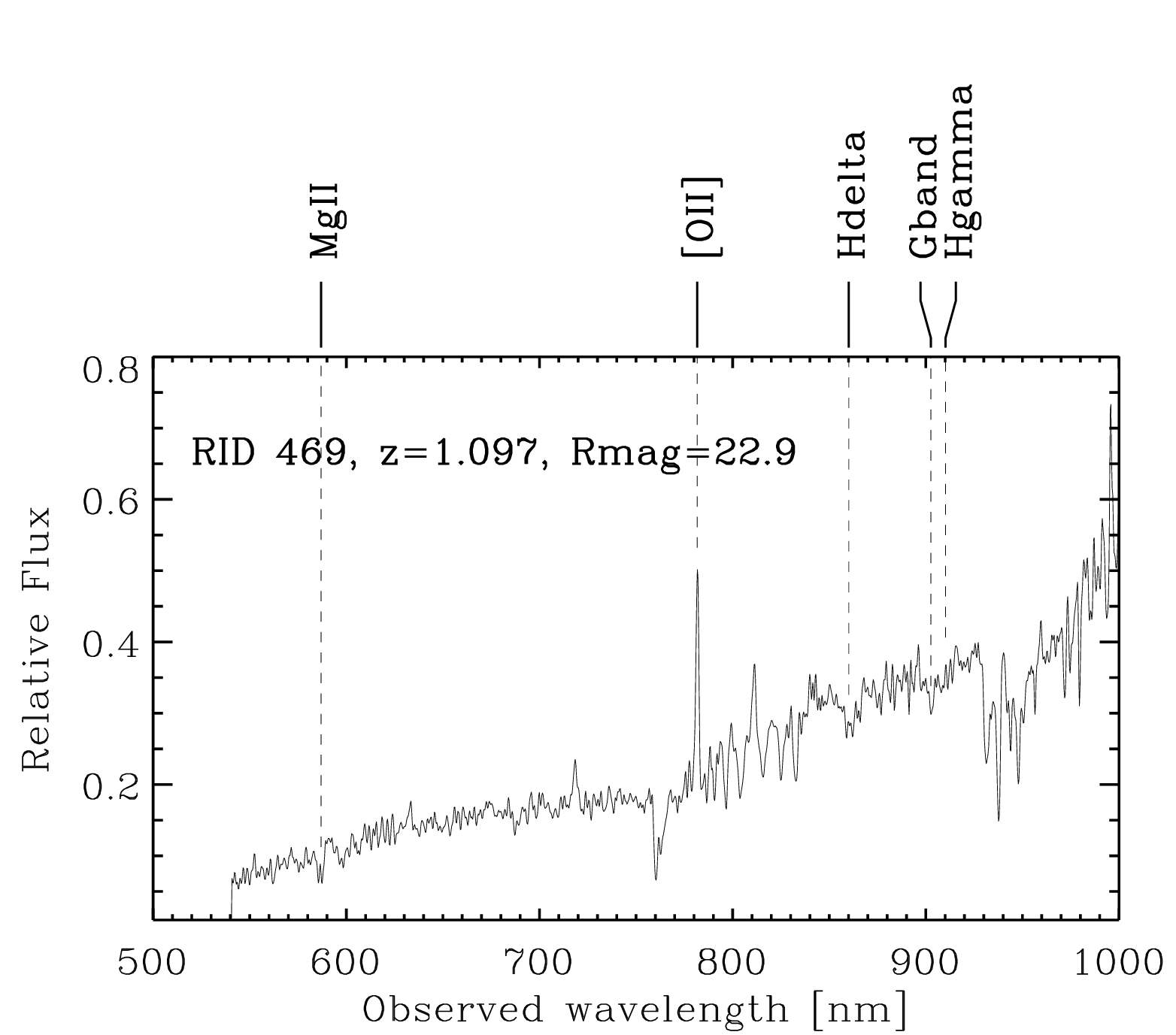} }}%
   \vspace{-5.0em}
\subfloat[RQ AGN]{{\includegraphics[width=8.0cm]{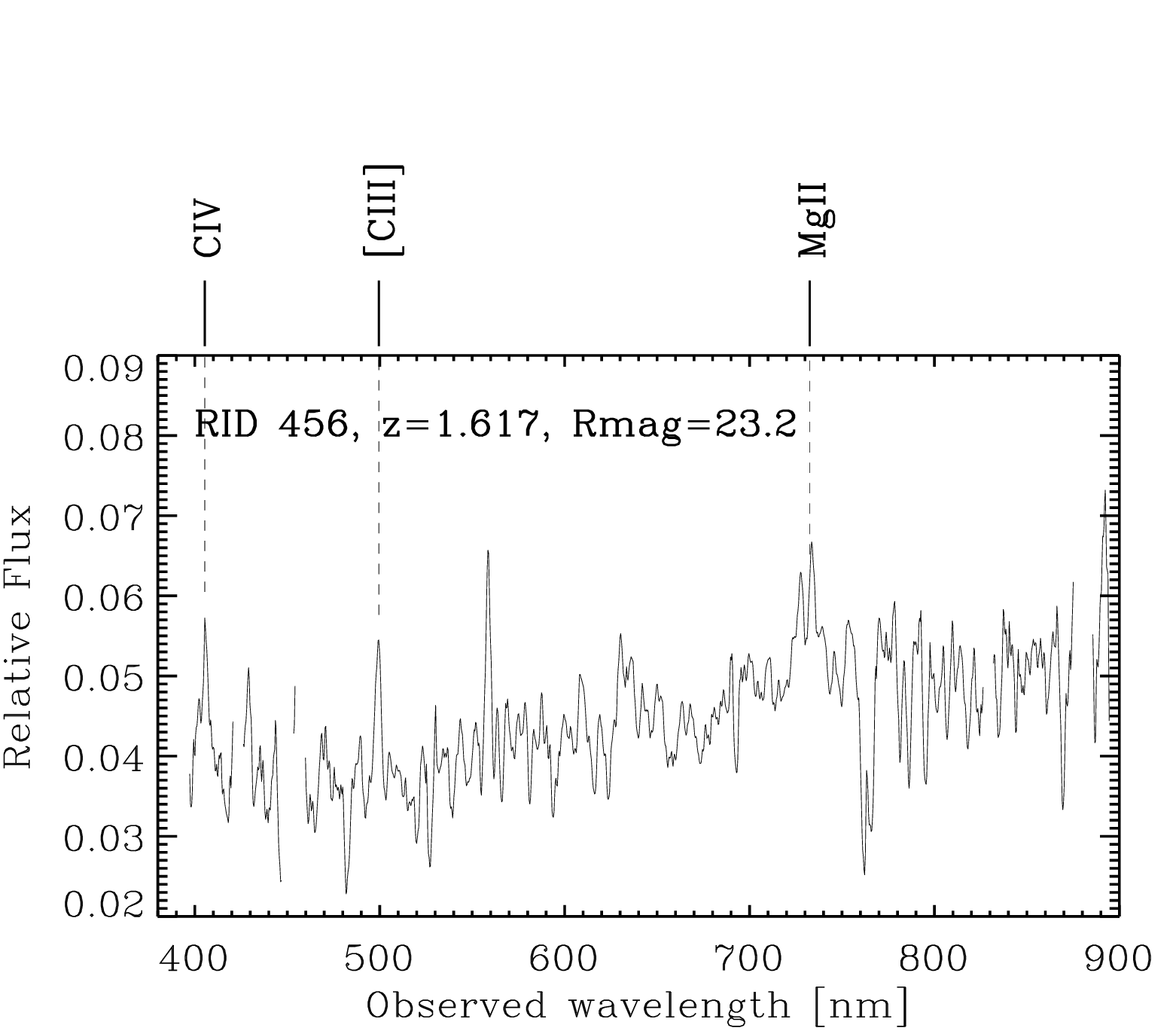} }}%
   \vspace{-3.5em}
\subfloat[RL AGN]{{\includegraphics[width=8.0cm]{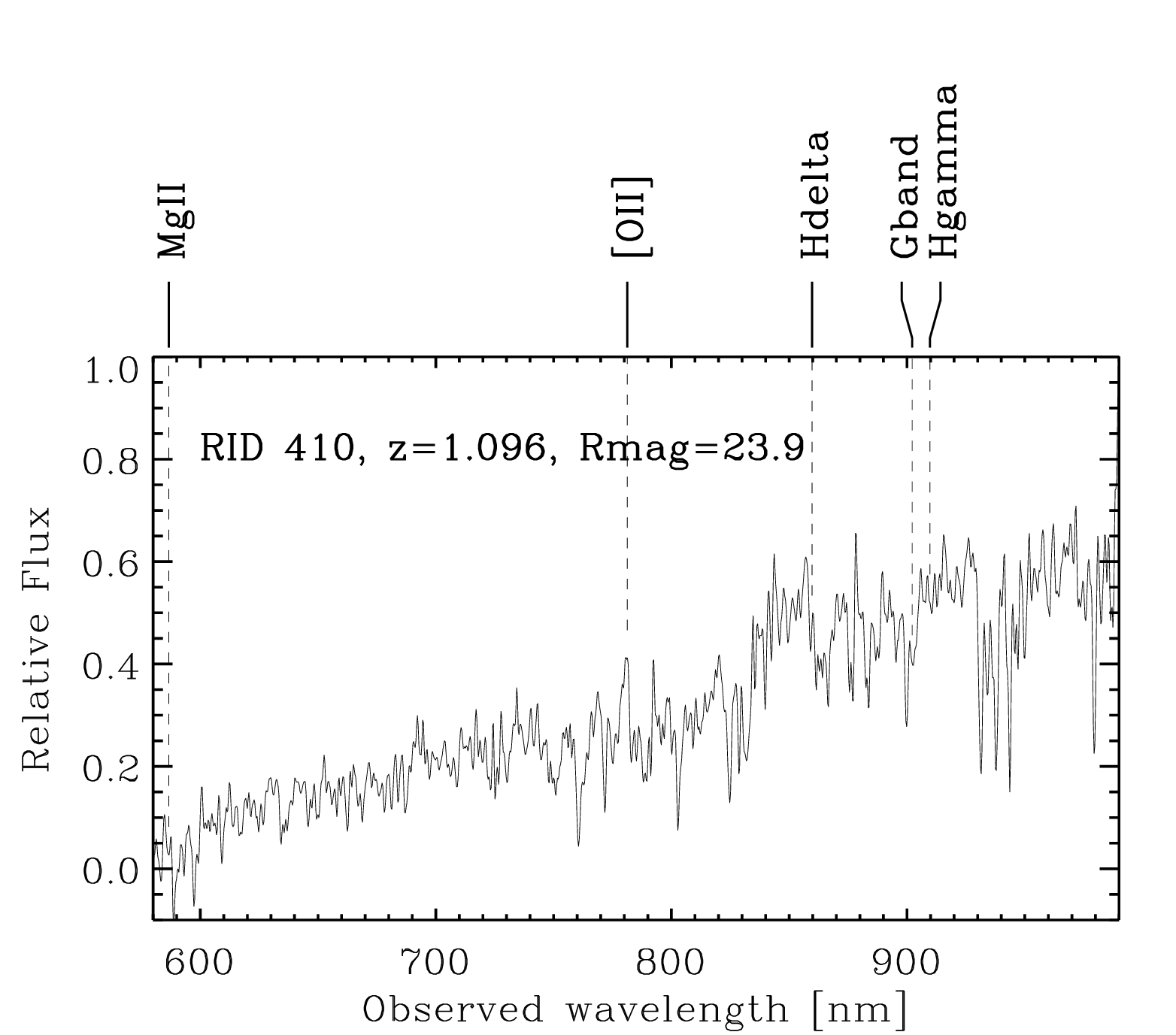} }}%
\caption{Examples of E-CDFS spectra of (from top to bottom) a SFG, a RQ,
  and a RL AGN having redshifts and magnitudes typical of the sample (and
  $44 \le S_{\rm 1.4GHz} \le 75~\mu$Jy). The SFG is bluer than the RL AGN
  but redder than the RQ one. The latter displays broad lines but its
  radio-quietness cannot be established from its spectrum \citep[see Table
    5 of][and references therein]{Bonzini_2012}.}
\label{fig:ECDFS_spectra}       
\end{figure}

In the past, when surveys were much shallower, optical counterparts were
relatively bright but telescopes were also smaller than they are today, so
one might think that the complexity of the problem has not changed
much\footnote{I though that for modern radio surveys things
  have become more complex than they used to be but after chatting with
  Robert Laing, who played a big role in the optical identification of the
  3CRR \citep{Laing_1983}, I am not so sure!}. To put things into
perspective, the median $R_{\rm mag}$ for the Extended {\it Chandra Deep
  Field}-South (E-CDFS) VLA sample, which reaches $S_{\rm 1.4GHz} \sim
32.5~\mu$Jy, is $\sim 23$; and this refers only to sources detected in the
$R$ band, as $\sim 20\%$ of the objects have only an IR counterpart
\citep{Bonzini_2012}. Getting spectra for such faint sources is very time consuming 
(prohibitively so for the very faint tail) but can in principle be done. 
Even if we had optical spectra for all the E-CDFS sources, though, 
this would not help us much as often for faint counterparts one can only
see a couple of lines. This is enough to get a redshift but not to properly
classify the object, as illustrated in Fig. \ref{fig:ECDFS_spectra}, which
shows examples of E-CDFS spectra of (from top to bottom) a SFG, a RQ, and
a RL AGN having redshifts and magnitudes typical of the sample. And
finally, optical based classification is well known to be prone to
obscuration biases (see below).
 
If one adds to all of the above the fact that the faint radio sky is a quite
heterogeneous mix (Sect. \ref{sec:components}), then the business of source
classification turns out to be quite a complex endeavour. I summarize here
the main indicators used to classify faint radio sources ranked in  
{\it rough} order of practical effectiveness\footnote{What follows below 
is an
  evolved and expanded version of Sect. 3 of \cite{Bonzini_2013}. The
  ranking order was mainly determined by looking at how many sources had
  their classification changed by a given indicator after the first three
  were applied to the E-CDFS sample.}. These stem from the properties of
radio sources as sketched in Sect. \ref{sec:components}. I stress that
indicators can be ranked low either because they are intrinsically weaker than
others or not as sensitive (at least for now).
\begin{enumerate}

\item {\bf FIR -- radio correlation}. This correlation has been discussed
  in Sects. \ref{sec:RQAGN} and \ref{sec:SFGs} and is usually parametrized
  through the so-called $q$ parameter, that is the logarithm of the FIR ($8
  - 1,000~\mu$m) to radio flux density \citep{Helou_1985}. Even with {\it
    Herschel}, which covers the $\sim 55 - 670~\mu$m range, one needs some
  extrapolation through templates to estimate the full FIR flux
  density. However, quite often the total FIR emission cannot be reliably
  derived because of lack of data at long wavelengths. Therefore $q_{\rm X}$ is
  used, where X can be, for example, $24~\mu$m or $70~\mu$m (the longer the
  wavelength, the better, to decrease the contribution from AGN heated dust
  [the torus]). Different papers have used the observed or rest-frame $q$
  to define a locus, or sometimes a dividing value, to differentiate
  between sources following the FIR -- radio correlation and those which do
  not. The latter display a ``radio excess'', which is characteristic of RL
  AGN. (Note that a RQ AGN with core radio flux density larger than the
  extended SF flux density will also have a slight radio excess:
  Sect. \ref{sec:emission_mech}. It therefore matters where one draws the
  line.)

\item {\bf X-ray power}. Only AGN can have hard X-ray power ($2 - 10$ keV)
  $L_{\rm x} \gtrsim 10^{42}$ erg s$^{-1}$ \cite[see][and references
  therein]{Szokoly_2004}. This does not mean that there are {\it no} AGN
  below this value, far from it: 1. $\sim 78\%$ of the RL AGN in
  \cite{Padovani_2015a} have $L_{\rm x} < 10^{42}$ erg s$^{-1}$; 2. and
  $\sim 6\%$ of RQ AGN with X-ray detection in the same paper are also
  below this cut. The first point is simply due to the fact that jet-mode
  AGN (which make up the majority of the E-CDFS RL sub-sample: 
  see Sect. \ref{sec:LF_RL})
  are not very strong X-ray emitters, while the second one is related to
  the (known) existence of low-power radiative mode AGN. Note that this
  indicator comes from the X-ray band, where (so far) the fraction of jet-mode AGN
  is small.

\item {\bf IRAC colour - colour diagram}. Different extragalactic sources
  occupy somewhat different regions of parameter space in {\it Spitzer}
  Infrared Array Camera (IRAC) colour - colour diagrams. One version of
  these plots the ratio $S_{8.0}/S_{4.5}$ versus $S_{5.8}/S_{3.6}$, where
  the flux densities refer to all four IRAC channels at 3.6, 4.5, 5.8, and
  $8.0~\mu$m \citep[][and references therein]{Donley_2012}. Quasars (RL and
  RQ) produce a power-law continuum across these bands, which makes them
  occupy a specific locus. The completeness of this selection method is
  therefore high ($\sim 75\%$) at $L_{\rm x} \ge 10^{44}$ erg s$^{-1}$ but
  relatively low ($\lesssim 20\%$) for $L_{\rm x} \le 10^{43}$ erg
  s$^{-1}$. IRAC selection appears also to be incomplete to RGs
  \citep{Donley_2012}. Nevertheless, mid-IR (MIR) selection is very
  important since it identifies also heavily obscured AGN, many of which
  are missed even by deep X-ray surveys \citep{Donley_2012}.

\item {\bf X-ray spectrum and variability}. Intrinsic X-ray absorption, the
  presence of a K-shell Fe line at 6.4 keV, and X-ray variability also
  discriminate between AGN and SFGs \cite[e.g.,][and references
    therein]{Vattakunnel_2012}. These indicators are of a higher order
  compared to the ``simple'' flux needed to derive $L_{\rm x}$ and
  therefore require a better S/N ratio.

\item {\bf Other radio indicators}. 

\subitem {\it Radio spectrum}. As discussed in Sect.
\ref{sed:quasars_blazars} and \ref{sec:SFGs}, SFGs (but also RGs) have
steep GHz radio spectra ($\alpha_{\rm r} \approx 0.7$), while flatter
spectra are typical of the compact core emission associated with AGN. An
inverted ($\alpha_{\rm r} < 0$) radio spectrum therefore excludes a SFG but
could be equally associated with a RL or RQ AGN.

\subitem {\it Radio morphology}. The presence of a bright, compact core or
clear radio jets/lobes suggests the presence of an AGN. Extended emission
on $\sim$ kpc scales with no obvious peaks or jets/lobes is more likely to
originate from SF, which however could also come from the host galaxy of a
RQ AGN. This indicator requires resolutions $\lesssim 1$ arcsecond
\citep[e.g.,][]{Richards_2007}.

\subitem {\it Radio power}. As amply discussed above, RGs and radio quasars
are more powerful than RQ AGN and SFGs. Nevertheless, while it might be safe to
assume that anything above $P_{\rm 1.4GHz} \approx 10^{24}$ W
Hz$^{-1}$ has nothing to do with SF, this is only valid at low redshifts
given the strong evolution of SFGs (see Fig. \ref{fig:flux_z} and
Sect. \ref{sec:evolution}). Additionally, such luminous sources are rare and
more normal, lower power radio sources can equally be associated with jet-mode
RGs, RQ AGN, or SFGs.

\item {\bf Optical indicators}.

\subitem {\it Optical spectra}. The presence of broad or high-exci\-tation
emission lines in optical spectra indicates AGN activity: after all, this
how quasars were discovered. Nevertheless, in modern radio surveys many sources
are too faint to get even a spectrum decent enough to determine a redshift. 
For example, only in $\sim 40\%$ of the sources with redshift information
in \cite{Padovani_2015a} is the redshift spectroscopic, 
being photometric in the remaining ones.

\subitem {\it BPT diagrams}. These use emission line ratios to distinguish
galaxies dominated by various photoionization processes, in particular to
separate SFGs from AGN \citep[they are named after the three authors
  of][]{Baldwin_1981}. While until recently these could only be applied at
$z \lesssim 0.3$, \cite{Kewley_2013} has presented a new diagnostic, which
can be used up to $z \sim 3.5$. Good enough optical spectra are however
still required.

As a rule, optical-only indicators are then useful solely for relatively
bright sources and therefore at relatively low redshifts. However, it is
also important to remember that the optical band does not give the full
story, being strongly affected by absorption and/or dilution by the host
galaxy. For example, there are many cases of sources, which show no sign
of nuclear activity in their optical spectra but are strong ($L_{\rm x}
\gtrsim 10^{43}$ erg s$^{-1}$) X-ray sources \cite[e.g.,][and references
  therein]{Smith_2014}.  And the identification of low accretion ($L/L_{\rm
  Edd} \lesssim 0.01$) AGN will be heavily affected by the properties of
their host \citep{Hopkins_2009b}.

\item {\bf VLBI detection}. VLBI detections of high-luminosity radio cores
  ($ \gg 2 \times 10^{21}$ W Hz$^{-1}$) are almost certainly AGN related,
  whereas lower-luminosity cores may be caused either by AGN or by
  supernova activity \citep[][and references
    therein]{Middelberg_2011}. This is a clear cut AGN indicator but its
  application is hampered by the still limited sensitivity of VLBI
  observations and for the same reason until recently worked mostly for RL
  AGN \citep[][but see Sect. \ref{sec:emission_mech}]{Bonzini_2013}.
  
\item {\bf Radio polarization}. Polarized GHz radio emission is supposed to
  originate mostly from the jets or lobes of extended AGN\footnote{Blazar
    cores are also polarized, but blazars are relatively rare in the faint
    radio sky.}, where coherent large scale magnetic fields are likely to
  be present \citep[e.g.,][and references therein]{Hales_2014b}. As such,
  this indicator should single out RGs. Indeed, the number counts of RL AGN
  in \cite{Padovani_2015a} are fully consistent (that is, somewhat above)
  the surface density of polarized sources derived by \cite{Hales_2014b},
  assuming a typical fractional polarization of $4\%$. Again, this
  indicator is of a higher order compared to simply measuring a radio flux
  density, so polarization observations are less sensitive.
  
\item {\bf $\bf D_{4000} - P_{\rm 1.4GHz}/M_{\star}$
  plane}. \cite{Best_2005a} used the location of sources on the $D_{4000} -
  P_{\rm 1.4GHz}/M_{\star}$ plane, where $D_{4000}$ is the strength of the
  4,000 \AA~break (a proxy for the mean stellar age of a galaxy) to
  separate SFGs from RL AGN. As for the FIR -- radio correlation, this
  method singles out ``radio excess'' sources, that is RL AGN. However, at
  variance with the former, it works only at low redshifts (or requires IR
  spectra at high redshifts), needs detailed modelling to get the stellar
  mass, and suffers from the relatively large uncertainties associated with
  the $D_{4000}$ technique \citep{Ellison_2016}.

\item {\bf $\bf R$ value}. The radio-to-optical flux density ratio has been
  proposed as an indicator of radio-loudness by \cite{Schmidt_1970} and a
  value $\sim 10$ has been long used since the seminal paper by
  \cite{Kellermann_1989}. As discussed by \cite{Padovani_2011b}, this
  definition is totally insufficient to identify RQ AGN when dealing with a
  sample, which includes also SFGs and RGs, as both classes are or can be
  (respectively) characterized by low $R$ values \citep[see Fig. 4
    of][]{Bonzini_2013}. $R$ was in fact defined for quasar (broad-lined)
  samples, where it could be assumed that the optical flux was related to
  the accretion disk, but loses its meaning as an indicator of jet strength
  if the optical band is dominated by the host galaxy, as is the case for
  jet-mode RGs. This indicator is mentioned here for historical reasons but
  does not have much value for the classification of faint radio sources
  (although a high R does indicate a RL AGN).

\end{enumerate}

In short, to classify faint radio sources one first selects RL AGN using a
variant of the IR -- radio correlation, then separates the RQ AGN from SFGs
using $L_{\rm x}$. The IRAC diagram is then used to recover (RQ) AGN missed
by the X-ray criterion. Finally, other indicators are applied to catch
possible outliers\footnote{I have made this sound easy but reality is, as
  usual, more complex. The reader should consult Sect. 3 of
  \cite{Bonzini_2013} to get a feeling for the many subtleties and possible
  complications. In particular, not all radio sources are X-ray detected
  even in the deepest fields.}. \cite{Padovani_2011b} have also shown
that, by applying mainly the first three criteria discussed here to
representative, well-known local sources, the correct classification is always
recovered.

Table \ref{tab:class} summarizes the role of all indicators, where a "Y"
indicates the class(es) for which the relevant indicator is useful and a
"$\sim$Y" denotes ``limited'' applicability (e.g., the optical indicator
for AGN). SF sources include objects dominated by SF processes in the band
under consideration (e.g., those following the FIR -- radio correlation).

\begin{table*}
\centering 
\caption{Effectiveness of faint radio source classifiers.}
\label{tab:class}       
\begin{tabular}{|  l  |  l  | r |  c | }  \hline
{\bf Indicator} &  \multicolumn{2}{c|}{\bf AGN} & {\bf SF sources} \\ \hline
               & {\bf RL}~~ & ~~{\bf RQ}  &  \\ \hline
FIR - radio correlation & #Y &   &  #Y \\
$L_{\rm x}$  &  $\sim$Y   & Y~  &  \\
IRAC diagram & \multicolumn{2}{c|}{$\sim$Y} & \\
X-ray spectrum and variability & \multicolumn{2}{c|}{#Y} &  \\
other radio & \multicolumn{2}{c|}{#Y} &  \\
optical & \multicolumn{2}{c|}{$\sim$Y} & $\sim$Y \\
VLBI detection & \multicolumn{2}{c|}{#Y} & \\
radio polarization & #Y &   &   \\
$D_{4000} - P_{\rm 1.4GHz}/M_{\star}$
  plane& $\sim$Y &  &  $\sim$Y \\ \hline
\end{tabular}
\end{table*}

So far I implicitly assumed that objects can be classified one way or the
other. I often see the term ``hybrid sources'' used in the literature to
describe objects having both AGN and SFG features. I find this term
somewhat confusing: RQ AGN are often hosted in SF galaxies and therefore
black hole and SF related processes are both going to play a role in these
sources. Nevertheless, the black hole is the prime driver, which is what we should be
interested in. So a RQ AGN with its radio emission produced by supernova
remnants is still an AGN, not a ``hybrid''.

I have spent quite some time on the classification of faint radio sources
for two reasons: 1. there appears to be some confusion in the literature on
this topic; 2. it is extremely relevant for the classification of the even
fainter radio sources, which will be detected in the near future
(Sect. \ref{sec:class_SKA}).

\subsection{The importance of selection effects}\label{sec:selection_effects}

Even if the classification problem is complex, it appears nevertheless to
be solvable. However, not all indicators discussed above have been
available since 1984. And those which were, gave only a biased view.

Optical classification of a sample always starts with the brightest
sources, which are obviously easier to observe. This gives rise to a strong
selection effect: as shown by Fig. 1 of \cite{Padovani_2009}, an optical
magnitude cut in a radio flux-limited sample produces a bias against
sources with large $R$ values, i.e., RL AGN. The fraction of blue, SF
sources, therefore, appears artificially increased. Only $\sim 44\%$ of the
93 radio sources forming a complete sample in \cite{Windhorst_1985} could
be identified through optical imaging and photometry, resulting in two
Galactic stars, 10 quasars, and 29 galaxies, most of them of the blue (SF)
type. The paper rightly stressed that ``it should be remembered that the
nature of the {\it unidentified} sub-mJy radio sources is unknown as yet''
but apparently this warning went unheeded. A series of subsequent papers,
in fact, perpetuated the SFG mantra. \cite{Rowan-Robinson_1993}, for
example, concluded that faint radio counts ($S_{\rm 1.4GHz} \ge 0.1$ mJy)
were dominated by starburst galaxies, on the basis of a sample for which
only $\sim 20\%$ of the radio sources had optical identifications.
\cite{Gruppioni_1999}, on the other hand, with a fraction of optical
identifications close to $50 \%$, deduced that $\sim 44\%$ of radio sources
with $0.2 \le S_{\rm 1.4GHz} \le 1$ mJy were instead early-type galaxies.

Another problem was the fact that the sensitivity of deep radio surveys
usually decreases with the distance from the centre of the field of
view\footnote{Nowadays the images corresponding to a few individual
  pointings are typically combined to form a mosaic image so this is less
  of an issue \citep[e.g.,][]{Miller_2013}. Still, it needs to be taken
  into account.}.  This means that the evaluation of the true fraction of
sources of a given class needs to take that into account by weighing
appropriately each object by the inverse of the area accessible at the flux
density of the source \cite[e.g.,][i.e., a source whose flux density could
  be reached only in $10\%$ of the area is worth ten sources, which could
  instead be detected over the full survey]{Padovani_2007}. Neglecting this
correction can obviously lead to wrong values if the population mix changes
with flux density, as indeed observed.

A more important, but subtler, cause of misinterpretation is the (simple)
fact that radio-based diagnostics will give radio-only information! 
An AGN in which most of the radio emission is related to SF in its host
galaxy, as discussed in Sect. \ref{sec:RQAGN}, will be classified as a SF,
neglecting the rest (and dominant part) of its emitted power, which
ultimately comes from black hole related processes. Only by including
other, and broader, classification criteria, as detailed above, can one
paint the full picture of a radio source. Which leads me to the most
important of all selection effects: the lack of multi-wavelength data!

Imagine in fact to be an astronomer in the late 1980s -- early 1990s. The
Infrared Astronomical Satellite (IRAS), launched in 1983, which provided
the first high sensitivity all-sky map at 12, 25, 60 and 100 $\mu$m, reached
 $\approx 200$ mJy at $25~\mu m$ \citep{Moshir_1992}. For $q_{\rm
  24\mu m} \sim 1.26$ \citep{Sargent_2010}, a SFG with $S_{\rm 1.4GHz} \sim
1$ mJy has a MIR flux $f_{\rm 24\mu m} \approx 18$ mJy, i.e., a
factor of 10 smaller. One could not then use the IR -- radio correlation
indicator. Only when the Infrared Space Observatory (ISO) was launched in
1995 could these flux densities be reached and even surpassed
\citep[e.g.,][]{Gruppioni_2003}. The {\it Spitzer} satellite, launched in
2003, which reaches $f_{\rm 24\mu m} \approx 40~\mu$Jy
\citep{Bethermin_2010} (on very small areas), can actually detect SFGs all
the way down to $S_{\rm 1.4GHz} \approx 2~\mu$Jy. And the first IRAC 
colour--colour cuts for AGN selection started to appear only in 2004 -- 2005
\citep[e.g.,][]{Lacy_2004,Hatziminaoglou_2005}.

\begin{figure*}
\centering
\includegraphics[width=10cm]{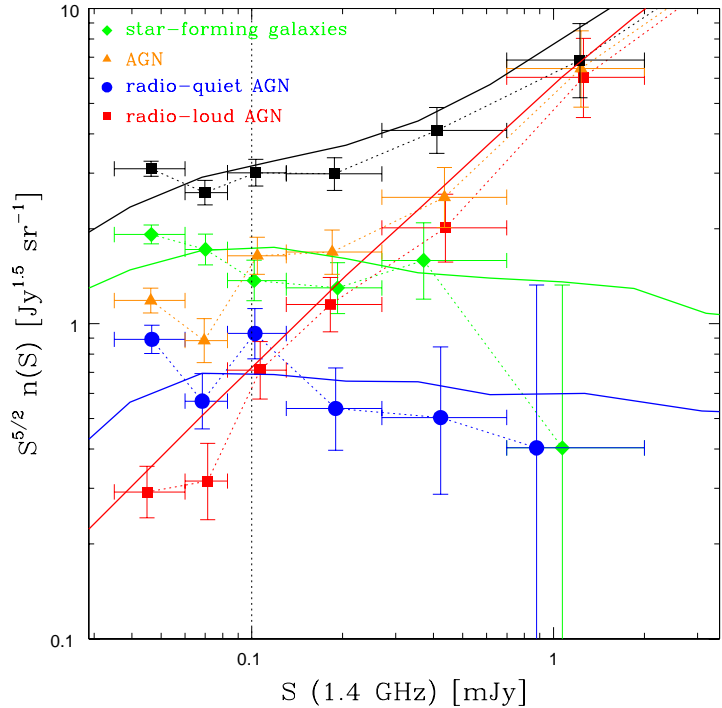}
\caption{Euclidean normalized 1.4 GHz source counts for the whole
  E-CDFS sample (black filled squares) and the various sub-classes of radio
  sources: SFGs (green diamonds), all AGN (orange triangles), RQ AGN (blue
  circles), and RL AGN (red squares). The solid lines are the SKADS
  simulated number counts \citep[][]{Wilman_2008}. Error bars correspond to $1\sigma$ Poisson
  errors \citep{Gehrels_1986}. The vertical dotted line marks the 0.1 mJy
  flux density. Adapted from \cite{Padovani_2015a}.}
\label{fig:counts_ECDFS}       
\end{figure*}

The story is roughly the same for the X-ray band. Before the launch of the
{\it Chandra} and {\it XMM-Newton} satellites in 1999 the best one could do with ROSAT
was a soft X-ray limit $f_{\rm x} \approx 10^{-15}$ erg cm$^{-2}$ s $^{-1}$
in {\it one} very small area of the sky \citep[$\sim 0.3$
  deg$^2$:][]{Lehmann_2001}. A RQ AGN with $S_{\rm 1.4GHz} \sim 1$ mJy has
$f_{\rm x} \approx 5 \times 10^{-15}$ erg cm$^{-2}$ s $^{-1}$
\citep{Padovani_2011a}, so the application of the $L_{\rm x}$ criterion was
possible only for the nine radio/X-ray sources in that area, seven of which
indeed had sub-mJy (5 GHz) flux densities \citep{Ciliegi_2003}. The first
results from the deep {\it Chandra} and {\it XMM-Newton} surveys appeared in 2001
\citep{Giacconi_2001,Hasinger_2001}. They reached a 0.5 - 2 keV flux $\sim
2 - 3 \times 10^{-16}$ erg cm$^{-2}$ s $^{-1}$, opening the way to their
use for the identification of faint radio sources \citep[and much
  more:][]{Brandt_2015}.

\subsection{Radio number counts by population}\label{sec:counts_by_pop}

Armed with the classifiers I have described, we can now ``solve'' the
question of the detailed composition of the sub-mJy radio sky.

\begin{figure}
\centering \includegraphics[width=8.4cm]{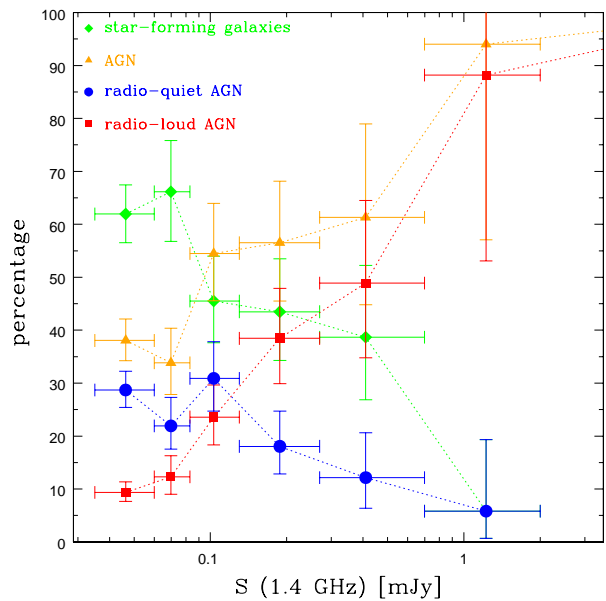}
\caption{Relative fractions (in percentage) of the various sub-classes of
  E-CDFS radio sources as a function of flux density: SFGs (green
  diamonds), all AGN (orange triangles), RQ AGN (blue circles), and RL AGN
  (red squares). Error bars correspond to $1\sigma$ Poisson errors
  \citep{Gehrels_1986}.}
\label{fig:fraction_ECDFS}       
\end{figure}

Figure \ref{fig:counts_ECDFS} presents the Euclidean normalized number counts
for the various sub-classes of the E-CDFS sample\footnote{I stress that I have 
not picked the E-CDFS sample to put together
Fig. \ref{fig:counts_ECDFS} and \ref{fig:fraction_ECDFS} (and
Fig. \ref{fig:ECDFS_local_LFs}) because this is my own survey. After
Sect. \ref{sec:RQAGN} and Fig. \ref{fig:flux_z}, in fact, one should expect
the presence of RQ AGN in radio surveys but the only radio counts I know of which include
{\it all} classes of astrophysical sources, which make up the faint radio
sky are those published by our group
\citep{Padovani_2009,Padovani_2011b,Padovani_2015a}. Having a deep,
sizeable radio sample, which is almost completely identified and where most
of the sources have a redshift (spectroscopic or photometric) is not
easy. But the real reason, I think, has to do with X-ray data: as shown by
Table \ref{tab:class} the best indicator of RQ AGN is X-ray power (i.e.,
once RL AGN are singled out using the FIR -- radio correlation, RQ ones are
easily identified through their $L_{\rm x}$). For that one needs very
deep X-ray data in a region of the sky where there are very deep radio data
as well: and this means the E-CDFS, which at present reaches
$f_{\rm 0.5
  - 2 keV} \sim 5 \times 10^{-18}$ erg cm$^{-2}$ s $^{-1}$
\citep{Lehmer_2012}. Note
  that we still do not detect {\it all} radio sources in the X-ray band but
  $\sim 60\%$ in the central region \citep{Vattakunnel_2012}.} \citep[full details
  in][]{Padovani_2015a}. These show the familiar steep slope followed by a
flattening below 1 mJy. Having classified the sources, however, one can go
beyond this well known behaviour, and see which classes are doing what. The
three main features of this figure are: 1. the fast drop of AGN, due to 
RL sources; 2. the rise of SFGs; 3. the rise of RQ AGN. These features are
better appreciated in Fig. \ref{fig:fraction_ECDFS}, which shows the
relative fractions of the different E-CDFS sub-classes. AGN go from being
totally dominant at large flux densities ($\gtrsim 1$ mJy) to being a
minority ($\sim 40\%$), and a very small one ($\sim 10\%$) as regards RL
sources. SFGs, on the other hand, are negligible at high flux densities but
become the dominant population below $\lesssim 0.1$ mJy, reaching $\sim
60\%$ at the survey limit. AGN make up $43\%$ of sub-mJy sources, which
shows that there are still plenty of them in the faint radio sky, while
SFGs represent $57\%$. RQ AGN constitute $26 \%$ of sub-mJy sources (or
$61\%$ of all AGN) but their fraction appears to increase at lower flux
densities, where they make up $75\%$ of all AGN and $\approx 29\%$ of all
sources at the survey limit, up from $\approx 6\%$ at $\approx 1$ mJy. So
the ``magical'' flux density is not 1 mJy but $\approx 0.1$ mJy, which is
where SFGs overtake AGN and also, for a strange coincidence, RQ AGN surpass
RL ones.

RQ AGN were very slow to appear in the deep radio field
arena. \cite{{King_2004}} and \cite{Jarvis_2004} were the first, to the
best of my knowledge, to include RQ AGN in the modelling of radio number
counts, while \cite{Simpson_2006} showed for the first time the existence
of RQ AGN in a deep ($S_{\rm 1.4GHz} \ge 0.1$ mJy) radio field, making the
suggestion that these sources (were a significant fraction of them very
absorbed) may dominate the population responsible for the flattening of the
normalized radio counts. The first observed radio number counts to include
RQ AGN were those of \cite{Padovani_2009}. 
Perhaps the ``radio-quiet'' name
has (unconsciously) fooled researchers for many years into thinking that
they were really ``radio silent''?

\subsection{Luminosity functions}\label{sec:LF}

The determination of the LF requires a complete, flux density-limited
sample of sources, with redshift, at distances large enough that peculiar
velocities cancel out ($z \gtrsim 0.003$).  For relatively small power
ranges one can consider a single power law of the type $\Phi(P) \propto
P^{-\epsilon}$, while for broader ranges a two power-law LF $\Phi(P)
\propto 1/[(P/P_*)^{\epsilon_1} + (P/P_*)^{\epsilon_2}]$ with a break at
$P_*$ might give a better fit. More complex models (e.g., a Schechter LF)
are of course also possible and often better.

All LFs in the Universe have $\epsilon > 0$, that is more powerful sources
are rarer than less powerful ones. The detection of the former, therefore,
requires large sampling volumes, which can be more easily obtained through
relatively shallow but wide area surveys. These, however, penalize
intrinsically faint objects, which are more easily detected by going deep,
which can be done effectively only on small regions of the sky (for
example, at a given redshift a lower flux density implies a lower radio
power; or alternatively, at a given radio power a lower flux density
implies a larger redshift: Fig. \ref{fig:flux_z}). Because of these two
conflicting requirements, many extragalactic surveys now follow the
so-called ``wedding cake'' strategy, in which multiple surveys are made
covering different depths and areas (with the two being inversely
proportional). In the radio band the NVSS, for example, covered the whole
sky at 1.4 GHz north of $-40^{\circ}$ declination down to $\sim 2.5$ mJy
\citep{Condon_1998}, while a series of deeper surveys have been carried out
on much smaller areas (Fig.  \ref{fig:counts_obs}).

Figure \ref{fig:counts_ECDFS} and \ref{fig:fraction_ECDFS} demonstrate
clearly the changing mix of radio sources as a function of flux density. As
a result, because of deeper surveys, the focus of the LF derivation has shifted 
greatly in recent years from what it used to be. In a radio sky
dominated by RGs, radio quasars, and blazars the emphasis was on flat
vs. steep sources, FSRQs vs. FR IIs, BL Lacs vs. FR Is \citep[][and
  references therein]{Urry_1995,deZotti_2010}. Figure 1 of
\cite{Padovani_2011a} shows the relatively small surface densities most of
these classes are predicted to have once one enters the sub-mJy regime. For
example, only $\approx 10$ blazars are expected in the E-CDFS area (0.285
deg$^2$), out of 765 sources. Nowadays the focus is on SFGs, RL, and
RQ AGN.

\begin{figure*}
\centering
\includegraphics[width=10cm]{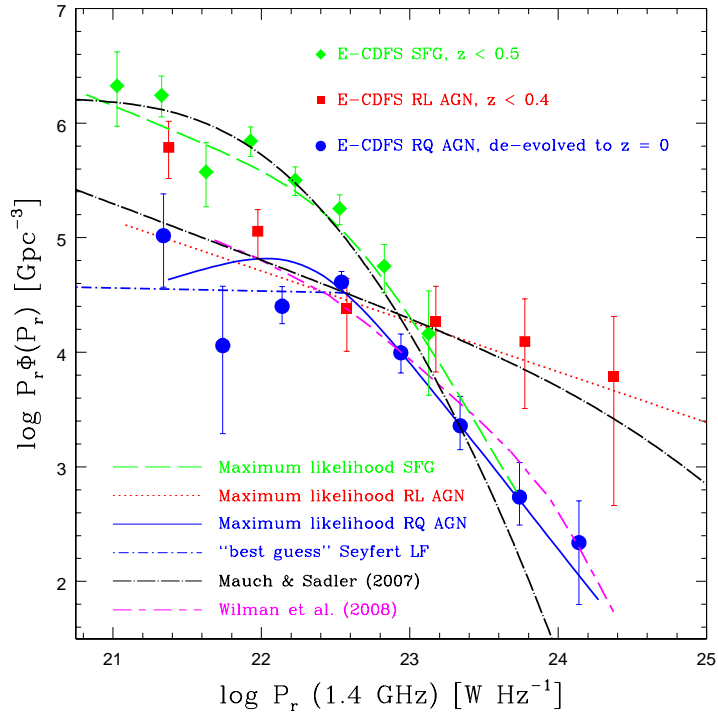}
\caption{The local differential 1.4 GHz LFs in a $P \times \phi(P)$ form
  for E-CDFS SFGs ($z < 0.5$; green diamonds), RL AGN ($z < 0.4$; red
  squares), and RQ AGN (whole sample de-evolved to $z=0$ using the
  1/$V_{\rm a}$ method and the best fit evolutionary parameter from a
  maximum likelihood fit; blue circles). The best-fit local LFs from a
  maximum likelihood fit are also shown, together with my ``best guess'' of
  the radio LF of Seyfert galaxies for $P_{\rm 1.4GHz} \lesssim 3 \times
  10^{22}$ W Hz$^{-1}$, the $z < 0.3$ LFs from \cite{Mauch_2007} for SFGs
  and AGN (upper and lower curve respectively), and the local LF for RQ AGN
  {\it assumed} in the SKADS simulation of \cite{Wilman_2008}. Error bars
  correspond to $1\sigma$ Poisson errors added in quadrature with the
  cosmic variance uncertainties for SFG and RL AGN and with the variations
  of the number density associated with a $1\sigma$ change in the
  evolutionary parameter for RQ AGN.
  }
\label{fig:ECDFS_local_LFs}       
\end{figure*}

Figure \ref{fig:ECDFS_local_LFs} shows the local differential 1.4 GHz LFs
\citep[derived using the 1/$V_{\rm a}$ method, a variation of the 1/$V_{\rm
    max}$ one: see][for details]{Schmidt_1968,Padovani_2015a} for E-CDFS
SFGs ($z < 0.5$; green diamonds), RL AGN ($z < 0.4$; red squares), and RQ
AGN (whole sample de-evolved to $z=0$ using the best fit evolutionary
parameter from a maximum likelihood fit; blue circles). The best-fit local
LFs from maximum likelihood fits are also shown \citep[][and Padovani et
  al., in preparation]{Padovani_2015a}. These are somewhat model dependent
but by making maximal use of the data provide additional information on the
low redshift LF, which is based on relatively small samples (plus they are
free from arbitrary binning). Finally, the figure displays also the $z <
0.3$ LFs from \cite{Mauch_2007} for SFGs and AGN and the local LF for RQ
AGN {\it assumed} in the SKADS simulation of \cite{Wilman_2008}
(Sect. \ref{sec:predictions}). The LFs are shown in a $P \times \Phi(P)$
form, which is almost equivalent to the $\phi(M_{\rm B})$ form\footnote{$P
  \times \Phi(P) = 2.5/ln(10) \times \Phi(M) \sim 1.09 \times \Phi(M)$,
  where the units of $\Phi(M)$ are mag$^{-1}$ volume$^{-1}$. Note that
  these units are also sometimes used in the radio band: e.g.,
  \cite{Condon_1989,Sadler_2002,Mauch_2007}. The conversion, instead, to
  units of Mpc$^{-3}$ dex$^{-1}$ used, for example, by \cite{Simpson_2012},
  is done by dividing my values by $10^9/{\rm ln(10)}$. The $P \times
  \Phi(P)$ form allows an easy separation of luminosity and density
  evolution as the former simply translates the LF to higher powers with no
  change in number, while the opposite is true for the latter.} normally
used in the optical band. I now discuss the (large amount of) information
conveyed by Fig.  \ref{fig:ECDFS_local_LFs} taking every population in turn
first. 
  
\subsubsection{SFGs}\label{sec:LF_SFG}

The local LF for SFGs is best fitted by two power laws with $\Phi(P)
\propto P^{-1.5^{+0.1}_{-0.2}}$ and $\propto P^{-3.2\pm0.2}$,
respectively below and above $P_* \sim 4 \times 10^{22}$ W Hz$^{-1}$
(Padovani et al., in preparation). The E-CDFS derivation is consistent
with previous ones based on sub-mJy data \citep[e.g.,][not shown for
  clarity]{Smolcic_2009a,Padovani_2011b,Mao_2012} but reaches $\sim 3
- 10$ lower powers because of its fainter flux density limit. It
also agrees impressively well with the local ($z < 0.3$) LF from the
sample of \cite{Mauch_2007}, which is based on a correlation of the
NVSS with galaxies brighter than $K = 12.75$ mag in the second
incremental data release of the 6 degree Field Galaxy Survey
(6dFGS). This is no mean feat: the NVSS/6dFGS SFG sample includes 4006
sources, all of them classified using high quality optical spectra. In
contrast, the E-CDFS SFG sample includes 91 ($z < 0.5$) or 356 (full
sample) sources, classified following the scheme described in
Sect. \ref{sec:deep_pop}. This agreement, therefore, validates {\it on
  a statistical basis} the classification scheme applied to the
E-CDFS sources. 

\subsubsection{RL AGN}\label{sec:LF_RL}
  
The local LF for RL AGN is best fitted by a single power law with $\Phi(P)
\propto P^{-1.44\pm0.4}$. The comparison with previous determinations based
on sub-mJy samples has to be done carefully because many papers make a
distinction between SFGs and AGN but not between RQ and RL AGN. The local
LF for {\it all} E-CDFS AGN compares well, for example, with that of
\cite{Smolcic_2009b} and \cite{Mao_2012}, while that of RL AGN agrees with
that of \cite{Simpson_2012} (once a correction is applied to take into
account their different definition of RL AGN). The agreement with the
NVSS/6dFGS LF \citep[and also with that of][]{Heckman_2014} is less good 
and the E-CDFS LF is a factor $\sim 2.2$ higher.
This is likely to be due partly to small number statistics 
but mostly to the somewhat different selection criteria. Many of the
samples put together by cross-correlating very large surveys, for example
those discussed by \cite{Heckman_2014}, by construction do not sample the
whole RL AGN population. They are, in fact, limited to galaxies (and
therefore do not include non-stellar and broad-line sources), bivariate
(i.e., the result of a cross match between a radio and an optical/NIR
survey), and often also restricted to steep-spectrum ($\alpha_{\rm r} >
0.5$) radio sources. The E-CDFS LF, being simply radio flux density
limited, has no such biases and therefore is bound to find larger number
densities; cf. the \cite{Mao_2012} LF, which is also a factor $\sim 3$
above the NVSS/6dFGS LF.

Finally, all RL AGN LFs from sub-mJy samples are well above ($\sim 2$
orders of magnitude) the radiative-mode radio AGN LF of \cite{Best_2014}
and their $L/L_{\rm Edd}$ are mostly $\lesssim 0.01$ \citep[see, e.g.,
  Figs. 6 and 12 of][respectively]{Padovani_2015a}. This shows that the
bulk of faint RL AGN are of the jet-mode type. I stress that the number
densities of ``classical'' (i.e., high flux density) RL quasars are orders of magnitude smaller than
those of the RL AGN in Fig.  \ref{fig:ECDFS_local_LFs} \citep[and
  off-scale: $\lesssim 10$ Gpc$^{-3}$ for $P \gtrsim 10^{26}$ W
  Hz$^{-1}$ for FSRQs; see, e.g.,][]{Wall_2005,Padovani_2007}.
  
\subsubsection{RQ AGN}\label{sec:LF_RQ}
  
The E-CDFS RQ AGN $z \le 0.4$ sample includes only 6 objects and the
corresponding LF is therefore very uncertain. I then use the LF for the
full sample de-evolved to $z=0$ using the best fit evolutionary parameter
from the maximum likelihood analysis and the local LF derived from the
latter.  The local LF for RQ AGN is best fitted by two power laws with
$\Phi(P) \propto P^{-0.6^{+0.6}_{-0.7}}$ and $\propto
P^{-2.6^{+0.2}_{-0.3}}$, respectively below and above $P_* \sim 2 \times
10^{22}$ W Hz$^{-1}$.  As discussed in \cite{Padovani_2015a} (see their
Fig. 8), this LF was found not be inconsistent with the radio LFs for three
samples of Seyfert galaxies (within the rather large errors). One can
actually use these LFs to estimate the RQ AGN LF at low radio powers
($P_{\rm 1.4GHz} \lesssim 3 \times 10^{22}$ W Hz$^{-1}$; with the caveat
that, being these bivariate, it is almost certain that the samples are not
complete in the radio and therefore the LFs are robust lower limits to the
true ones), which is what I have done in Fig. \ref{fig:ECDFS_local_LFs}
(dot-dashed blue line).

No comparison can be made with previous determinations based on
radio-selected samples because none exists. I can compare the
derived RQ AGN LF with that {\it assumed} in the SKADS simulation (magenta
short-long-dashed line: see Sect. \ref{sec:predictions}). This is based on a simple conversion of the AGN
X-ray LF to a radio LF using a linear relationship between X-ray and radio
power \citep[see details in][]{Wilman_2008}. Given the assumptions behind
the latter, the agreement between the two is surprisingly good and strongly
suggests that: 1. both the SKADS approach and the classification scheme
applied to the E-CDFS sources are validated; 2. the sources we are
selecting in the radio band are the same as the (RQ) AGN selected in the X-ray band;
3. the ``best guess'' Seyfert LF is indeed likely to be a lower limit.

\subsubsection{SFGs and AGN}

The main messages of Fig. \ref{fig:ECDFS_local_LFs} come out loud only when
the three LFs are considered together. Namely: 1. faint RL AGN have a much
flatter radio LF than RQ ones and are always predominant, especially so at
$P_{\rm 1.4GHz} \gtrsim 3 \times 10^{22}$ W~Hz$^{-1}$; 2. the RQ AGN LF
appears to be somewhat parallel to, and overlapping with, the SFG one at
$P_{\rm 1.4GHz} \gtrsim 5 \times 10^{22}$ W~Hz$^{-1}$; 3. AGN dominate over
SFG at $P_{\rm 1.4GHz} \gtrsim10^{23}$ W~Hz$^{-1}$, in agreement with
previous studies \citep[e.g.,][]{Mauch_2007}.

\subsection{Evolution}\label{sec:evolution}

AGN and SFGs evolve strongly
\citep[e.g.,][respectively]{Merloni_2013,Gruppioni_2013}, i.e., their LF
changes with redshift, with their powers and/or they numbers being
different from what they are at $z \sim 0$. This was quickly realised for
quasars thanks to the first radio surveys and for radio sources in general
even before the discovery of quasars (Sect. \ref{sec:obs_counts}).  A
classic (but not the first) paper is \cite{Schmidt_1968}, who showed that
the space density of 3CR quasars at $z = 1$ was $\approx 100$ times
the local density. Furthermore, to
reproduce the relatively narrow hump in the observed Euclidean normalized
counts (Fig. \ref{fig:counts_obs}), \cite{Longair_1966} suggested that not
all radio sources evolved equally but that only the most powerful sources
could evolve strongly with redshift. Many more papers followed: for
example, \cite{Dunlop_1990}, by mostly using a sample with $S_{\rm 2.7GHz}
\ge 100$ mJy, pointed out that this strong evolution could not continue to
very high redshifts but a decrease in the number density of FSRQs, SSRQs,
and RGs had to take place around $z \sim 2 - 4$.

This topic is very vast. As is the case for LFs (Sect. \ref{sec:LF}), its
focus has recently shifted as we reach fainter radio flux densities. I
refer to \cite{deZotti_2010} for a comprehensive review of the evolution of
strong radio sources. Here I only mention the fact that there appears to be
a consensus on the strong and positive evolution of FSRQs/SSRQs/FR IIs
(i.e., radiative-mode RL AGN), and on the weak, if any, evolution of BL
Lacs/FR Is (i.e., jet-mode RL AGN) \citep[see also][and references
  therein]{Padovani_2007,Giommi_2012}. I now concentrate on the most recent
results pertinent to the faint radio sky, taking every sub-population in
turn. Before I do that let us have another look at
Fig. \ref{fig:counts_ECDFS} keeping in mind what we have learnt from
Fig. \ref{fig:counts_exa}: SFGs have normalized counts, which rise at lower
flux densities, as is the case for RQ AGN, while RL AGN display the opposite
behaviour. This already tells us, without any doubt and need for further
information, that the first two sub-samples are undergoing positive
evolution, at variance with RL AGN.

\subsubsection{SFGs}\label{sec:evol_SFG}

E-CDFS SFGs evolve strongly, with $P(z) \propto (1+z)^{2.1\pm0.1}$
for $0 < z \le 3.25$; there is also evidence that the strength of
the evolution decreases with redshift (Padovani et al., in
preparation). Furthermore, \cite{Padovani_2011b} found suggestive evidence
of such a slowing down in the CDFS, with $P(z) \propto
(1+z)^{3.5^{+0.4}_{-0.7}}$ and $\propto (1+z)^{1.6^{+0.6}_{-0.7}}$ for $z
\le 1.3$ and $1.3 < z \le 2.3$ respectively.

Until recently, no {\it direct} determination of the radio evolution of sub-mJy
SFGs was possible, likely for lack of redshifts. For example,
\cite{Hopkins_2004} combined the constraints from the global (radio to
X-ray) SFR density evolution with those derived from the 1.4 GHz sub-mJy
source counts to infer $P(z) \propto (1+z)^{2.7\pm0.6}$ and $\phi(P)
\propto (1+z)^{0.15\pm0.60}$ imposing a redshift cutoff at $z=2$ (i.e.,
basically a PLE). One of the first direct estimates was that of
\cite{Smolcic_2009a}, who derived $P(z) \propto (1+z)^{2.1-2.5\pm0.2}$
for $z \le 1.3$. This is weaker than found by \cite{Padovani_2011b}, although not
significantly so but, as discussed in \cite{Bonzini_2013}, the method employed by 
\cite{Smolcic_2009a} to
separate AGN and SFGs, based on rest-frame optical colours, is not optimal.
\cite{McAlpine_2013} obtained $P(z) \propto (1+z)^{2.5\pm0.1}$ for $z \le
2.5$ but their source classification is based only on UV to K band
photometry, with their AGN being selected as sources redder than the spiral
galaxy templates (i.e., having early type hosts) and SFGs being everything else
(which groups at least some of the RQ AGN with the SFGs).

\subsubsection{RL AGN}\label{sec:evol_RL}

RL AGN are the only sources, which are present in significant numbers both
in the bright and faint radio skies. High-power radiative mode objects, however, 
are more common at large flux densities while low-power jet mode ones are 
predominant at lower ones (as per Fig. \ref{fig:flux_z}). Because of the
extra complications in selecting RQ AGN (Sect.  \ref{sec:class}),
many papers dealing with the evolution of sub-mJy radio
sources have unfortunately lumped RQ and RL AGN together, obtaining an overall relatively
weak evolution \citep[e.g.,][]{Smolcic_2009b,McAlpine_2013}, which 
masked the big difference between the two AGN sub-classes.
\cite{Padovani_2011b} were the first to study them separately in the CDFS
sample and found that RL AGN evolve (in number) strongly but negatively
($\phi(P) \propto (1+z)^{-3.7^{+1.1}_{-1.6}}$), while RQ AGN evolve (in
power) strongly but positively ($P(z) \propto (1+z)^{2.5^{+0.4}_{-0.5}}$).

\cite{Padovani_2015a} have exploited the better statistics of the E-CDFS
sample and found that the evolution of RL AGN is still a PDE but a complex
one: the number density evolves {\it positively} as $\phi(P) \propto
(1+z)^{2.2^{+1.8}_{-1.6}}$ up to $z_{\rm peak} = 0.5\pm0.1$, beyond which
it declines steeply $\propto (1+z)^{-3.9^{+0.7}_{-0.8}}$ (the large error
bars at low redshift reflect the small number of RL AGN with $z \le 0.5$).
This is consistent with other results derived at larger flux densities. 
\cite{Rigby_2015} studied various
samples of {\it steep-spectrum} ($\alpha_{\rm r} > 0.5$) AGN selected from
a variety of radio surveys with increasingly smaller areas and flux density
limits down to 0.1 mJy. They found that the number density peaks at a
luminosity-dependent $z_{\rm peak}$ for $P_{\rm 1.4GHz} \gtrsim 10^{26}$ W
Hz$^{-1}$, with the most powerful sources peaking at earlier times than the
weaker ones. Below this value $z_{\rm peak}$ appears to remain constant. At
their lowest powers, $P_{\rm 1.4GHz} \sim 2 \times 10^{24}$ W Hz$^{-1}$,
$z_{\rm peak} = 1.1\pm0.4$ (or possibly $1.1^{+0.2}_{-1.1}$).
Since the median $P_{\rm 1.4GHz}$ of E-CDFS RL AGN is $\sim 10^{24}$ W Hz$^{-1}$ a
$z_{\rm peak}$ $\sim 0.5$ is in agreement with their results. The work by
\cite{Rigby_2015} puts on stronger footing previous results on the
high-redshift decline of strong radio sources \citep[see][]{deZotti_2010}.

\cite{Best_2014} have also studied samples of (steep-spectrum)
radio AGN selected from a variety of surveys down to 0.2 mJy and up to
$z=1$. They classify their sources into radiative-mode and jet-mode AGN
using emission line diagnostics. The space density of the jet-mode
population with $P_{\rm 1.4GHz} \lesssim 10^{24}$ W Hz$^{-1}$ stays
constant up to $z \approx 0.5$ and then decreases; at moderate
powers, $10^{24} \lesssim P_{\rm 1.4GHz} \lesssim 10^{26}$ W Hz$^{-1}$, the
space density increases to $z \sim 0.5$ before falling. At the highest
powers the space density appears to increase up to $z \sim 1$ but the
statistics is somewhat limited (see their Fig. 5). Based on Sect.
\ref{sec:LF_RL}, the large majority of E-CDFS RL AGN are of the latter
type, so the E-CDFS results are similar, taking into account the somewhat
more limited coverage of the luminosity -- redshift plane. 
 \cite{Simpson_2012} also found zero or negative evolution for RL AGN with
 $P_{\rm 1.4GHz} \lesssim 10^{24}$ W Hz$^{-1}$ (although their RL AGN
 selection is somewhat different: Sect. \ref{sec:LF_RL}). As for
 radiative-mode RL AGN, the evolution in their space density appears to be
 comparable to the evolution of RQ radiative-mode AGN selected in other
 bands \citep[e.g.,][]{Best_2014}.

\subsubsection{RQ AGN}

The evolution of RQ AGN in the radio band has been first determined by
\cite{Padovani_2011b} and more recently updated by \cite{Padovani_2015a}
(no other determinations exist to the best of my knowledge). The E-CDFS RQ
sample is consistent with a PLE of the type $\propto (1+z)^{2.5\pm0.2}$ for
$0 \lesssim z < 3.7$, not very different from that of SFGs.  As was the
case for SFGs, however, there is ($2 \sigma$) evidence of a slowing down at
higher redshifts, with $P(z) \propto (1+z)^{4.0\pm0.6}$ and $\propto
(1+z)^{2.0\pm0.5}$ for $z \le 1.3$ and $1.3 < z \le 3.7$
respectively. Radio selected RQ AGN share the strong evolution of the
powerful, radiative-mode RL AGN but have radio powers more similar to those
of the jet-mode ones, which make up most of the sub-mJy RL AGN
(Fig. \ref{fig:ECDFS_local_LFs}), a situation which can be somewhat
confusing.

\section{The big picture: what does it all mean?}

One obvious question, as this point, is: what are we learning by studying
the faint radio sky? In particular, what do the LF and evolution of faint
radio sources tell us of astrophysical relevance? I address this next by
providing very specific examples.

\subsection{Astrophysics of faint radio sources}\label{sec:astrop}

\subsubsection{SFGs and cosmic star formation history}\label{sec:astrop_SFG}

``Once, there were no stars. [...] Understanding how [...] gas evolved into
the Universe filled with stars that we observe today [...] remains one of
the most important goals of modern astrophysics'' \citep{Mac_2013}.  The
solution to this puzzle is related to the time dependence of the SFR
density (SFRD, generally expressed in units of $\rm M_{\odot}$ yr$^{-1}$
Mpc$^{-3}$), the so-called ``Lilly-Madau plot''. This appears to have peaked
$\approx 3.5$ Gyr after the Big Bang and to have declined
exponentially since then \citep[][and references therein]{Madau_2014}. As mentioned in
Sect. \ref{sec:SFGs}, radio emission in SFGs trace the SFR, so radio
observations can also be relevant. Traditionally, however, their role has
been somewhat limited \citep[but see,
  e.g.,][]{Haarsma_2000,Seymour_2008,Karim_2011} although other, more
``standard'' methods all have their own limitations.  One issue is that
even the deepest radio data at present can only reach relatively high SFR
at high redshifts: e.g., $\approx 1000~\rm M_{\odot}$ yr$^{-1}$ at $z \sim
3$ \citep[][Fig. 1a]{Madau_2014}.

One way one can derive the SFRD is through the power density, defined as
$\rho_{\rm L} = \int P \Phi(P) dP$. Parametrizing the LF as in
eq. \ref{eq:evol}, one obtains the redshift dependence of $\rho_{\rm L}$ as
follows:

\begin{equation}
\rho_{\rm L}(z) = f_{\rm D}(z) f_{\rm L}(z) \int P_0 \Phi(P_0) dP_0 =
\rho_{\rm L}(0) f_{\rm D}(z) f_{\rm L}(z),
\label{eq:rho_L}
\end{equation}

where $\Phi(P_0)$ is the local LF. In other words, the power density at
redshift $z$ is the power density at $z=0$ multiplied by both luminosity
and density evolutionary functions (assuming, of course, these have no
dependence on power). If $f_{\rm D}(z) = (1+z)^{k_D}$ and $f_{\rm L}(z) =
(1+z)^{k_L}$ then $\rho_{\rm L}(z) \propto (1+z)^{k_D + k_L}$ \citep[see
  also][]{Hopkins_2004}. The power density can then be transformed into a
SFRD by using the relevant SFR --  power conversion, whose robustness
depends on the band at hand. For example, FIR emission is straightforward
to understand in the optically thick case for an intensely SFG: $\rm
SFR_{\rm FIR}~[M_{\odot}$ yr$^{-1}] = 4.5 \times 10^{-44} \rm~L_{\rm FIR}$
[erg s$^{-1}$] \citep{Kennicutt_1998}. UV and optical indicators, on the
other hand, are extremely sensitive to dust, while radio emission is more
indirect, since it relies on the complex and not fully understood physics
of cosmic-ray generation and confinement
\citep[e.g.,][]{Condon_1992,Bell_2003}. As a result, the exact conversion
factor of the latter is still debated. \cite{Bell_2003} gives $\rm SFR_{\rm
  radio}~[M_{\odot}$ yr$^{-1}] = 5.5 \times 10^{-22} \rm~P_{\rm 1.4GHz}$ [W
  Hz$^{-1}$] (for $P_{\rm 1.4GHz} \ge 6.4 \times 10^{21}$ W Hz$^{-1}$:
below this value the relationship is slightly non-linear). The
existence of the FIR -- radio correlation (Sect. \ref{sec:SFGs}) is in 
any case a very strong argument for using radio power as a SFR proxy.

For a {\it linear conversion} between SFR and power, therefore, SFRD$(z)
\propto f_{\rm D}(z) f_{\rm L}(z) \propto (1+z)^{k_D + k_L}$, i.e., the
observed slope of the SFRD$(z)$ relationship constraints $k_D + k_L$. Said
differently, when one measures the evolution of the LF of SFGs one is also
constraining the evolution of the SFRD in the Universe. \cite{Madau_2014}
found SFRD$(z) \propto (1+z)^{2.7}$ for $z \ll 1.9$ and $\propto
(1+z)^{-2.9}$ for $z \gg 1.9$, with a smooth transition in between (see
their eq.  15)\footnote{They note that a solid interpretation of the time
  dependence of the SFRD from first principles is still missing
  \citep[e.g.,][]{Mac_2013}.}. The low redshift behaviour is not
inconsistent with our current understanding of the radio evolution of SFGs
(Sect. \ref{sec:evol_SFG}). Constraining the high redshift evolution is
tougher, though, as shown (again) by Fig. \ref{fig:flux_z}: even the E-CDFS
can detect at $z \gtrsim 2$ only SFGs with $P_{\rm 1.4GHz} \gtrsim 6 \times
10^{23}$ W Hz$^{-1}$, which correspond to the relatively high end of the LF
\citep[Fig. 2 of][shows that the fraction of E-CDFS SFGs at $z > 2$ is
  indeed quite small.]{Padovani_2015a}

\cite{Karim_2011} used a large 1.4 GHz survey of the COSMOS field and a 
{\it Spitzer} $3.6~\mu$m selected sample to carry out the most extensive
study in the radio band to date. Through stacking in bins of $\rm
M_{\star}$ and (photometric) redshift and converting their mean $S_{\rm r}$
to SFRs, they computed the integrated SFRD, finding a monotonic decline in
the SFRD from $z = 3$ to today. In other words, their results suggest a
peak in the SFRD at $z > 3$, at variance with \cite{Madau_2014}.  However,
as stated in their paper, their data cannot constrain the situation at high
redshifts as strongly as at $z < 1.5$ and therefore they cannot rule out
a SFRD peak at $1.5 < z < 3$. This topic is picked up again in
Sect. \ref{sec:future_SFG}.

\subsubsection{RL AGN and quiescent galaxies}\label{sec:astrop_RL}

AGN are well-known to evolve positively, that is they were more luminous
and/or more numerous at higher redshifts. Why is it then that faint RL AGN
appear to display the opposite behaviour? This can be understood by
looking at AGN evolution from a broader, modern perspective. Our
current understanding is that the number density of more luminous AGN
peaks at redshifts higher than those of lower luminosity objects (the so-called
downsizing). That is, sources in a given luminosity range increase in
number from lower to higher redshifts up to a maximum redshift, $z_{\rm
  peak}$, above which their numbers decrease, with $z_{\rm peak}$ 
strongly correlated with power. This behaviour has been seen at many
wavelengths   
\citep[e.g.,][and Sect. \ref{sec:evol_RL} (for the
  radio)]{Hopkins_2007,deZotti_2010,Merloni_2013}.

The E-CDFS can detect sources as weak as $P_{\rm 1.4GHz} \approx 6
\times 10^{23}$ W Hz$^{-1}$ and $\approx 2 \times 10^{24}$ W Hz$^{-1}$ up
to $z \sim 2$ and $\sim 3$, respectively (Fig. \ref{fig:flux_z}). For
$P_{\rm 1.4GHz} \lesssim 10^{26}$ W Hz$^{-1}$ $z_{\rm peak} \approx 1$
\citep{Rigby_2015}. This means that at the median power of RL AGN ($
P_{\rm 1.4GHz}\sim 10^{24}$ W Hz$^{-1}$) one can probe the evolutionary
behaviour of the RL population well beyond $z_{\rm peak}$ and already in
its declining phase. Hence the strong negative evolution.

So the real question becomes: what is driving this evolution?
\cite{Padovani_2011b} made, as far as I know, the first connection between
the negative evolution of RL AGN\footnote{The blazar community has been
  aware of the likely negative evolution of a sub-class of BL Lacs for
  quite some time: see \cite{Giommi_2012} and references therein.}  and
that of elliptical galaxies, by noticing the similarity between their
results and those of \cite{Taylor_2009}, who found that the number density
of $\rm M_{\star} > 10^{11} \rm M_{\odot}$ red galaxies declined as
$\Phi(z) \propto (1+z)^{-1.6}$ for $z \le 1.8$. \cite{Best_2014} took this
idea further by assuming that jet-mode (since we are dealing here with
low-power sources) RL AGN are hosted in quiescent (i.e., non SF) galaxies
and combining the known stellar mass function of the host galaxies with the
prevalence of jet-mode RL AGN as a function of $\rm M_{\star}$. They then
came up with a $\Phi(z) \propto (1+z)^{-0.1}$ out to $z=0.8$ and $\Phi(z)
\propto (1+z)^{-6.5}$ at higher redshifts \citep[][have modified the former
  into $\propto (1+z)$, which gives a better match to their
  data]{Rigby_2015}. The physical reasons behind this evolution are
complex, hotly debated, and not entirely sorted out but are related to SF
being ``quenched'' as time goes by, which translates into a decrease in the
number density of quiescent galaxies at higher redshifts
\citep[e.g.,][]{Peng_2012}\footnote{This is a very active field with many
  papers published on the subject in the past few years: see, e.g.,
  \cite{Madau_2014,Somerville_2015} and references therein.}.
 
As for the radiative-mode AGN (both in the radio and other bands), it has
been noted many times that the black hole mass growth rates derived from
the AGN bolometric LF (which evolves strongly and positively with redshift)
track closely the cosmic SFRD, which has led to the suggestion that SF and
black hole growth are linked. This would make sense, as both mechanisms are
fed by the gas in the host galaxy, albeit on quite different spatial
scales. Nevertheless, ``the differences between accretion histories published in
the recent literature would caution that it is premature to consider this
comparison to be definitive'' \citep{Madau_2014}.

\subsection{The origin of radio emission in (sub-mJy) RQ AGN}\label{sec:emission_mech}

I have mentioned in Sect. \ref{sec:RQAGN} that the mechanism responsible
for the {\it bulk} of radio emission in (non-local) RQ AGN has been a
matter of debate for the past fifty years or so \citep[see also the
  Introduction of][]{Condon_2013}. Alternatives have included a scaled down
version of the RL AGN mechanism \citep[e.g.,][]{Miller_1993,Ulvestad_2005},
perhaps because the central black hole is rotating more slowly than in RL
AGN \citep{Wilson_1995}, SF \citep{Sopp_1991}, coronal emission 
\citep[][and references therein]{Raginski_2016}, and more 
\citep[e.g.,][and references therein]{Orienti_2015}.

This is a highly non-trivial issue for various reasons: 1. most ($> 
90\%$) AGN are RQ; 2. some of the proposed explanations have profound
implications on our understanding of AGN physics (jets, accretion, black
hole spin, etc.); 3. some others are very relevant for the relationship
between AGN and star formation in the Universe and the co-evolution of
supermassive black holes and their host galaxies (related to ``AGN
feedback''), which is a very hot topic in extragalactic research
\citep[e.g.,][for recent reviews]{Kormendy_2013,Heckman_2014}.

\begin{figure}
\centering
\includegraphics[width=8.4cm]{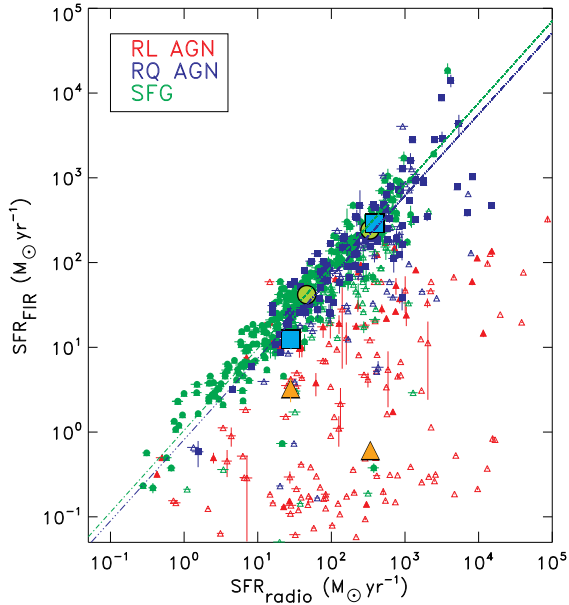}
\caption{SFR derived from the FIR luminosity versus the SFR from $P_{\rm
    1.4GHz}$ for the E-CDFS sample. SFGs are plotted as green circles, RQ
  AGN as blue squares, and RL AGN as red triangles. Full symbols represent
  sources detected in at least one Photoconductor Array Camera and Spectrometer 
  (PACS) filter, while sources shown as
  empty symbols are \textit{Herschel} non-detections, for which SFR$_{\rm
    FIR}$ is less robust.  Large symbols with lighter colours are the
  results of a stacking analysis. The two lines are the best fits for the
  SFGs and RQ AGN with PACS detection. Figure reproduced from
  \cite{Bonzini_2015}, Fig. 3, with permission.}
\label{fig:SFR_comparison}       
\end{figure}

The study of the faint radio sky can help here: for the first time one can
select RQ AGN, RL AGN, and SFGs in the radio band and within the same
sample. Sub-mJy RQ AGN share many properties with SFGs, including the
strong evolution, similar LF, and host galaxies, while not many with the
sub-mJy RL AGN. This might suggest that radio emission in RQ AGN is more
related to SF than to the central AGN.  One can directly test this by
comparing the SFR derived from the FIR luminosity as traced by {\it
  Herschel}, which is a very robust SFR estimator, with that estimated from
the radio power under the assumption that it is due to SF
(Sect. \ref{sec:astrop_SFG}). This is exactly what has been done by
\cite{Bonzini_2015} for the E-CDFS sample, as shown in Fig.
\ref{fig:SFR_comparison}.

The two SFR estimates for SFGs, not surprisingly, are in agreement over
four decades in SFR with a dispersion of 0.2 dex. More interestingly, there
is very good agreement between the SFRs derived from the two different
tracers also for RQ AGN with only a slightly larger scatter of 0.23
dex. This implies that the {\it main} contribution to radio emission in
{\it sub-mJy radio-selected RQ AGN at $z  \sim 1.5$} (the
mean redshift of the E-CDFS RQ AGN sample) comes from SF activity in the host
rather than from the black hole. This is further supported by the behaviour of
RL AGN, which populate the region below the best fit lines, since their
$\rm SFR_{\rm radio}$ is overestimated because of the strong jet/central AGN
contribution. Note that RQ AGN occupy the same locus as SFGs also in the
SFR - $\rm M_{\star}$ plane, suggesting that the majority of the host
galaxies of radio-selected RQ AGN are not significantly different from the
inactive galaxy population \citep{Bonzini_2015}\footnote{The E-CDFS RQ AGN
  have been selected mostly based on them falling within the SFG locus in
  the $q_{\rm 24\mu m} - z$ plane \citep{Bonzini_2013}. One could therefore
  argue that the correlation they follow in Fig. \ref{fig:SFR_comparison}
  is a consequence of the selection method. However, SFGs and RQ AGN have
  different MIR characteristics: SFGs, for example, have an average
  dispersion in $q_{\rm 24\mu m~obs} \sim 0.33$ dex, which is twice as
  small as that of RQ AGN. This is mainly due to the relatively large AGN
  contribution at MIR wavelengths in many RQ AGN. In other words, one can
  effectively use the MIR to discard RL AGN, but this band is not good
  enough to obtain a reliable estimate of the SFR, for which one needs the
  FIR (i.e., {\it Herschel}).}.

How can these results be reconciled with some of those reviewed in
Sect. \ref{sec:RQAGN}?  As discussed in \cite{Smolcic_2015}, at least part
of these differences might be attributed to the diversity of the
samples. \cite{Rosario_2013}, for example, have shown that in RQ,
relatively low-luminosity AGN, much of the observed radio emission is
consistent with SF in the AGN hosts, at variance with
\cite{Zakamska_2016}. Since their sources have IR powers $\nu L_{\nu}
(12\mu m) \lesssim 10^{44}$ erg s$^{-1}$ ($L_{\rm bol} \lesssim 10^{45}$
erg s$^{-1}$) one might think that there is a dependency on bolometric
power, which could be due to the likely different host galaxies, with
quasar-like sources being hosted in bulge-dominated galaxies and
Seyfert-like ones in disc-dominated galaxies
(Sect. \ref{sec:RQAGN}). Indeed, the E-CDFS RQ AGN are also of relatively
low power, having $\langle L_{\rm x} \rangle \sim 10^{43}$ erg s$^{-1}$
(i.e., $L_{\rm bol} \approx 3 \times 10^{44}$ erg s$^{-1}$). Nevertheless, this still
does not explain the \cite{Kimball_2011} and \cite{Condon_2013} results,
which refer to quasars.

Another complication might have to do with evolution
\citep{Padovani_2011b}: if the AGN related radio component is non-evolving,
as is the case for low-power RL AGN (Sect. \ref{sec:evol_RL}), while the SF
related one follows the evolution of SFGs (Sect. \ref{sec:evol_SFG}),
higher redshift RQ AGN should have their radio emission more SF dominated
than lower redshift ones. Both the \cite{Kimball_2011} and the
\cite{Zakamska_2016} samples, however, are at relatively low redshifts ($0.2 < z <
0.3$ and $z < 0.8$ respectively).

Further support for the SF connection in sub-mJy RQ AGN comes from high
resolution radio imaging. \cite{Richards_2007} have studied 92 radio
sources with $S_{\rm 1.4GHz} \ge 40~\mu$Jy in the Hubble Deep Field North
well resolved by MERLIN and the VLA at $0.2 - 2$ arcsec resolutions. They
found that the presence of an AGN is indicated in at least half of the 45
radio starbursts with X-ray counterparts. Furthermore, almost all extended
radio starbursts at $z > 1.3$ host X-ray selected obscured AGN (with
$L_{\rm x} < 10^{44}$ erg s$^{-1}$). These results are fully consistent
with a very close relationship between SF and radio emission in 
relatively high-redshift RQ AGN.

\cite{Chi_2013} have detected with VLBI 12 out of the 92 sources
studied by \cite{Richards_2007}. Of these, four fulfil the RQ AGN criteria
laid out in Sect. \ref{sec:class} \citep[based on $q_{\rm 24\mu m}$ and
  $L_{\rm x}$, the latter from][]{Richards_2007}, and have $0.7 < z < 4.4$
and $S_{\rm VLBI}/S_{\rm VLA} \gtrsim 0.5$.
This indicates that AGN emission makes up $> 50\%$ of the total in these
sources. Nevertheless, these authors estimate that 48/92 sources were bright
enough to be detected, which gives a detection rate of only 25\%. So the
majority of these sources have their arcsecond
scale emission completely resolved out by the VLBI beam, which suggests an
extended, possibly SF related source for the RQ AGN. 

Very recently \cite{Maini_2016} have also detected with VLBI compact
cores accounting for $\sim 50 - 70\%$ of the total radio emission in two
E-CDFS RQ AGN at $z \sim 1.4$ \citep[see also][who observed 18 COSMOS
  sub-mJy RQ AGN and detected three]{Herrera_2016}. Once this core emission
is removed both sources, which had a slight radio excess in Fig.
\ref{fig:SFR_comparison}, fall nicely on the FIR -- radio correlation.

In summary, some sub-mJy RQ AGN show evidence for relatively strong
compact radio cores, which suggests that the AGN component might be at the
same level as, or even stronger than, the SF one. One needs to
keep in mind, though, that VLBI detections might be biased towards
AGN-dominated sources, as at present they require relatively large flux
densities, and sizeable, complete, and fainter samples should be targeted
at VLBI resolutions to get a less biased picture. The {\it Herschel} results, 
moreover (Fig. \ref{fig:SFR_comparison}), point towards a dominance of SF    
in the radio emission of sub-mJy RQ AGN at $z \sim 1.5$.

\subsection{Radio emitting AGN in the larger context}\label{sec:AGN_context}

\begin{figure}
\centering
\includegraphics[width=8.4cm]{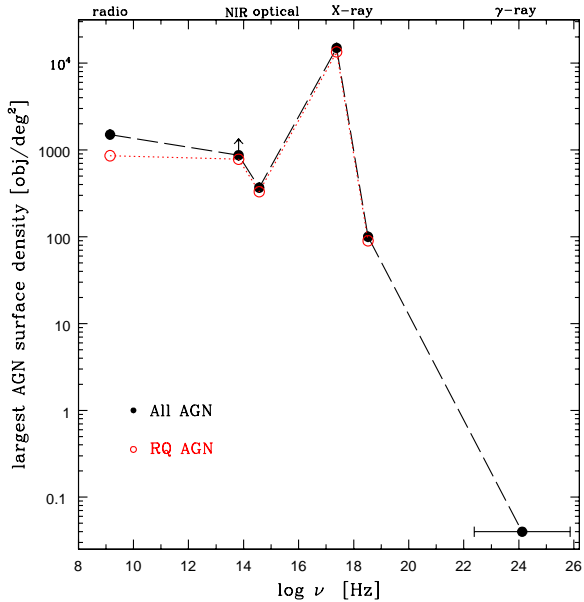}
\caption{The largest AGN surface density over the
  whole electromagnetic spectrum. Black filled points refer to all AGN,
  while open red points are for RQ AGN. The latter are actually measured
  only in the radio band, while in the NIR to X-ray bands they have been
  derived by simply subtracting $10\%$ from the total values. References
  for the relevant samples and facilities are: E-CDFS/VLA, radio \citep[1.4
    GHz:][]{Padovani_2015a}; COSMOS/{\it Spitzer}, NIR \citep[$4.5
    \mu$m:][]{Donley_2012}; VIMOS VVDS/VLT (Type 1) and zCOSMOS/VLT (Type
  2), optical \citep[I band:][respectively]{LeFevre_2013,Bongiorno_2010}; 4
  Ms/{\it Chandra}, soft X-ray \citep[$0.5 - 2$ keV:][]{Lehmer_2012}; {\it
    NuSTAR}, hard X-ray \citep[$8 - 24$ keV:][]{Harrison_2015}; 3FGL/{\it
    Fermi}, $\gamma$-ray \citep[100 MeV -- 300 GeV:][]{3FGL}. Based on the
  discussion in Sect. \ref{sec:class} the surface density in the NIR is
  only a lower limit to the true value since, by definition, IR selected
  AGN only account for those objects where the AGN dominates over the host
  galaxy at the wavelengths of interest, which imposes biases on the
  sample.}
\label{fig:agn_density}       
\end{figure}

After the discovery of quasars in 1963 AGN have been observed and detected
in all bands, which provide different windows on AGN physics. To put things
into perspective, Fig. \ref{fig:agn_density} shows my best estimates of the
largest surface density in various bands covering the whole electromagnetic
spectrum for all AGN (black filled points) and RQ AGN only (open red
points). There are no RQ AGN detected in the $\gamma$-rays\footnote{With
  the exception of NGC 1068 and NGC 4945, two Seyfert 2 galaxies in which
  the $\gamma$-ray emission is thought to be related to their starburst
  component \citep{Ackermann_2012b}.} \citep{Ackermann_2012a}.
   
Figure \ref{fig:agn_density} is a complex mix of physics, selection effects,
and technological limitations (perhaps providing enough material for
another review!). The main point I want to make is that the surface density
of radio-selected RQ AGN is already larger than that reached by the deepest
optical surveys and at the same level as the NIR values. The whole idea of
looking in an efficient way for RQ AGN in the radio band would have sounded
like an oxymoron until only a few years ago! The fact that this is now a
reality goes a long way to show how much radio astronomy has changed; and
this is just the beginning. The soft X-ray band wins the competition by
more than one order of magnitude as the current radio flux density limit is
still not as deep as the equivalent one in the X-rays \cite[see][and
  discussion therein]{Padovani_2015a}. However, it is important to remember that
radio observations are unaffected by absorption and therefore are sensitive
to all types of AGN, irrespective of their orientation (i.e., Type 1s and
Type 2s), unlike soft X-ray ones.
  
Figure \ref{fig:agn_density} can also be read as describing the ``detection
potential'' of the various bands. The actual number of {\it detected and
  identified} AGN is still heavily biased towards the optical/NIR bands, as
most of them were discovered through dedicated large-area spectroscopic
surveys. For example, of the 510,764 AGN in the Half Million
Quasars Catalogue \citep{Flesch_2015}, which includes mostly Type 1
sources, only $\sim 9\%$ and $\sim 11\%$ have been detected in the radio
and X-ray bands, respectively. Note that many more sources have been
detected (although often not identified) in the radio band.  For example,
the unified radio catalogue put together by
\cite{Kimball_2008,Kimball_2014}\footnote{\url{http://www.aoc.nrao.edu/~akimball/radiocat_2.0.shtml}}
by combining five radio catalogues (FIRST, NVSS, GB6, WENSS, and VLSSr) and
the SDSS includes almost three million sources north of $-40^{\circ}$. The
vast majority of these are going to be AGN, mostly RGs and RL quasars (and
there are more radio catalogues covering the sky south of $-40^{\circ}$).

\section{The future}\label{sec:future}

\subsection{New radio facilities}\label{sec:new_facilities}

Radio astronomy is at the verge of a revolution, which will usher in an era
of large area surveys reaching flux density limits well below current
ones. The Square Kilometre Array (SKA)\footnote{Everything there is to know
  about the SKA can be found at http://www.skatelescope.org. I give here
  only a very brief description of the project.}, in fact, will offer an
observing window between 50 MHz and 20 GHz extending well into the 
  nanoJy (nJy) regime with unprecedented versatility. Phase 1 (SKA1) will
constitute about 10\% of the full telescope and will take place between
2018 and 2023, with early science observations being conducted as early as
2020 with a partial array. The plan is for SKA1 to be followed by a Phase 2
(SKA2), which will complete the design, and should last until the late
2020s. Around 1 GHz, SKA1 will provide a major advance over existing
instruments. Resolution, sensitivity, and survey speed will be an order of
magnitude better in most cases, and in combination will occupy a new region
of performance.

The SKA will not be the only participant to this revolution. In 2011, a
decade long upgrade project has resulted in the VLA expanding greatly its
technical capacities and being renamed the ``Karl G. Jansky Very Large
Array''
(JVLA\footnote{http://science.nrao.edu/facilities/vla}). LOFAR\footnote{http://www.astron.nl/radio-observatory/astronomers/lofar-astronomers}
has started operations and is carrying out large area surveys at $15 - 200$
MHz \cite[e.g.,][]{Morganti_2010,vanWeeren_2014,Heald_2015,Williams_2016},
opening up a whole new region of parameter space at low radio
frequencies. The Murchison Widefield Array, a low-frequency radio
telescope operating between 80 and 300 MHz and one of three telescopes
designated as a precursor for the SKA, has also recently become operational
and is conducting surveys \citep{Hurley-Walker_2014}.

Many other radio telescopes are currently under construction in the lead-up
to the SKA, including
APERTIF\footnote{http://www.astron.nl/general/apertif/apertif} (The
Netherlands), ASKAP\footnote{http://www.atnf.csiro.au/projects/askap/}
(Australia), e-MERLIN\footnote{http://www.e-merlin.ac.uk} (UK), and
Meer\-kat\footnote{http://www.ska.ac.za/meerkat} (South Africa), with both
ASKAP and Meerkat being the other two of the three telescopes designated as
SKA precursors. All these projects will survey the sky vastly faster than
is possible with existing radio telescopes producing surveys covering large
areas of the sky down to fainter flux densities than presently available,
as fully detailed in \cite{Norris_2013}. \cite{Prandoni_2015} provide an
overview of the radio continuum surveys best suited to enable the top
science cases to be tackled by SKA1.

\subsection{How many sources, and of what type, will the new radio facilities detect?}

\subsubsection{Predictions}\label{sec:predictions}

One would like to have an idea of the number and type of sources likely to
be detected by these future deep surveys, for various reasons:
1. predicted surface densities allow one to plan ahead in terms of the
sheer number of sources expected; 2. the forecast on the type of sources is
important to plan the identification process and particularly to know in
advance, which kind of ancillary data will be more relevant.

The ``classic'' work in this respect is the SKADS
\citep{Wilman_2008,Wilman_2010}, a semi-empirical
simulation\footnote{The simulation is accessible at
  \url{http://s-cubed.physics.ox.ac.uk/s3_sex}.} of the extragalactic radio
continuum sky down to 10 nJy at 151, 610 MHz, 1.4, 4.86 and 18 GHz
including various source types: RQ AGN, RL AGN (RGs [FR Is and IIs], SSRQs,
FSRQs and blazars, and GPS), and SFGs (quiescent and starbursting). Sources
are drawn at random from the observed (or suitably extrapolated) radio
LFs. As clearly stated in the papers there are numerous
uncertainties and limitations in the SKADS simulations, as they had to rely,
by necessity, on extrapolations, being based on relatively high
flux density samples. This affects particularly the highest redshifts,
which can be better probed at fainter flux densities.
  
Figure \ref{fig:counts_obs} shows the SKADS simulated number counts 
(red dashed line) compared with smoothed versions of the observed
counts excluding (solid black line) and including (dotted black line) the
\cite{Owen_2008} sample. \citep[I consider here these two cases separately
  because the rapid rise observed in the counts for this sample is unique
  and might be caused by count corrections made for partial resolution of
  extended sources:][]{Condon_2012,Vernstrom_2016}. The agreement between simulations and
observations is very good down to 0.1 mJy, which is not surprising as
\cite{Wilman_2008} did compare their predictions with the data
down to this flux density.
  
Figure \ref{fig:counts_ECDFS} shows the SKADS simulated number counts 
 for all sources, SFGs, RL, and RQ AGN. The agreement with the E-CDFS
counts is impressively good, for the whole sample but also for all
sub-classes, even more so considering the scatter between various surveys
shown by Fig. \ref{fig:counts_obs}.
The simulated RL AGN Euclidean normalized counts are dominated (by a factor
$> 10$) by FR I-like sources, in agreement with the observed predominance
of jet-mode AGN. Based on the SKADS simulation, a deep survey reaching
$1~\mu$Jy at 1.4 GHz is expected to have a surface density of
$6.2 \times 10^4$ sources deg$^{-2}$. Application of eq. \ref{eq:counts}
using the LFs and evolution derived from the E-CDFS sub-samples gives
exactly the same result. The SKADS reaches 10 nJy, while
\cite{Padovani_2011a} gives {\it order of magnitude} surface densities down
to even smaller flux densities ($< 1$ nJy).

Another, very different approach to estimate the number of faint sources is
the so-called P(D) method \citep[or probability of
  deflection:][]{Scheuer_1957}, which uses background fluctuations to model
the source counts below the limit of a given survey \citep[e.g.,][and
  references therein]{Vernstrom_2014}. \cite{Vernstrom_2015} quote a value
of $8.9 \times 10^4$ sources deg$^{-2}$ for $S_{\rm 1.4GHz} \ge 1~\mu$Jy,
based on \cite{Vernstrom_2014}. The $\sim 40\%$ difference with the SKADS
results (visible also in Fig. \ref{fig:counts_obs}) is likely due 
to the (known) limitations of the SKADS simulation.

What type of sources will populate the faint radio sky?
Figs. \ref{fig:counts_ECDFS} and \ref{fig:fraction_ECDFS} already hint at
the answer: the fraction of SFGs is clearly on the rise, while that of AGN
is decreasing overall. This is quantified by the SKADS simulation, which
predicts that down to $S_{\rm 1.4GHz} = 1~\mu$Jy 80\% of the sources will
be SFGs, followed by RQ (12\%) and RL AGN (8\%). Quite similar fractions
($\sim 84\%$, 11\%, and 4\%) are obtained from the E-CDFS LFs and
evolution. To the best of our knowledge, therefore, a deep survey reaching
$\sim 1~\mu$Jy at $\sim 1$ GHz over 30 deg$^2$ will detect $\sim 4 \times
10^5$ ``potential'' AGN \citep{Smolcic_2015}, i.e., of the same order of
{\it all} currently known AGN (and candidates) included in the Million
Quasars catalogue\footnote{http://quasars.org/milliquas.htm.} and with a
surface density roughly equal to that of (current) X-ray selected AGN
(Fig. \ref{fig:agn_density}). The Evolutionary Map of the Universe (EMU),
one of the ASKAP surveys, expects to reach flux densities $\sim 2$ times
larger than those of the E-CDFS over $\sim 3/4$ of the sky, detecting $\sim
70$ million sources, about half of which will be ``potential" AGN
\citep{Norris_2011}. The ``potential'' here is important because, as
detailed above, the classification of faint radio sources requires a great
deal of ancillary, multi-wavelength information, which will not be easy to
get at very faint levels or over very large areas, as I am going to discuss
now.

\subsubsection{Source classification in the era of the SKA and its precursors}\label{sec:class_SKA}

It took more than thirty years to figure out the source population of the
$\lesssim 1$ mJy radio sky because source classification was complex but
above all because the relevant, and necessary, data at other wavelengths
were not available (Sect. \ref{sec:class} and \ref{sec:selection_effects}). 
What will the situation be for
the $\lesssim 1~\mu$Jy (GHz) sky? To answer this question, based on
Sect. \ref{sec:class}, one needs first to have at least order of magnitude
estimates of the X-ray, optical/NIR, and MIR\,--\,FIR fluxes sub-$\mu$Jy
radio sources are likely to have (where the X-ray, MIR\,--\,FIR, and
optical/NIR data are needed for source classification and
photometric/spectroscopic redshifts respectively). The current and, above
all, future availability of the relevant multi-wavelength data needs then
to be evaluated\footnote{I performed both tasks in \cite{Padovani_2011a},
  whose main results I summarise and update here. Needless to say, the
  sensitivities of future facilities are inherently uncertain, especially
  if the latter have not yet been approved for construction.}.

In the X-ray band, sources with $S_{\rm 1.4GHz} \sim 1~\mu$Jy should have
$f_{\rm 0.5 - 2 keV} \approx 10^{-17}$, $\approx
10^{-18}$, and well below $10^{-18}$ erg cm$^{-2}$
s$^{-1}$ for RQ AGN, SFGs, and RL AGN respectively. The deepest X-ray
survey currently available is the 4 Ms CDFS , which reaches $f_{\rm 0.5 - 2
  keV} \sim 5 \times 10^{-18}$ erg cm$^{-2}$ s$^{-1}$ over $\sim 0.1$
deg$^2$ \citep{Lehmer_2012}; the results of further 3 Ms of data should be
available soon (Luo et al., in preparation). The {\it Athena} mission,
selected by the European Space Agency (ESA) as the L2 mission (due for
launch in 2028), will reach $f_{\rm 0.5 - 2 keV} \sim 2 \times 10^{-17}$
erg cm$^{-2}$ s$^{-1}$ in 1 Ms \citep{Barcons_2015} but given its survey
speed will be able to cover larger areas much more efficiently than {\it
  Chandra}. At these levels, however, {\it Athena} is not only background
but also confusion limited and integrating further will not improve the
sensitivity (A. Rau, private communication). This means that even {\it
  Athena} surveys will not detect the bulk of the $\gtrsim 1~\mu$Jy
population (which will likely be made up of SFGs). Below this flux density,
very few, if any, radio sources will have an X-ray counterpart in the
foreseeable future, as there is no X-ray mission, existent or planned,
capable of detecting them.

As regards the MIR\,--\,FIR bands {\it Spitzer}, by reaching $f_{\rm 24\mu
  m} \approx 40~\mu$Jy in the GOODS fields, can detect now SFGs down to
$S_{\rm 1.4GHz} \approx 2~\mu$Jy (Sect. \ref{sec:selection_effects}). The
Space Infrared Telescope for Cosmology and Astrophysics (SPICA), which is
under consideration as a medium-class mission under the framework of the
ESA Cosmic Vision with a target launch in the mid-2020s, will have a
$24~\mu$m continuum sensitivity $\sim 10 - 50~\mu$Jy (1 hr, $5\sigma$) and
in the FIR band will improve upon {\it Herschel} by almost two orders of
magnitude \citep{Nakagawa_2015}. The bulk of the $\gtrsim 1~\mu$Jy
population should then be easily detected in the MIR (and perhaps FIR)
band, while deep SPICA exposures should be able to detect many radio
sources at $S_{\rm 1.4GHz} \lesssim 0.1~\mu$Jy.

What does this mean in practice for source classification? Identification
of RL AGN should be possible down to $S_{\rm 1.4GHz} \approx 0.1~\mu$Jy,
but these sources are likely to constitute only a minority ($\lesssim 5 \%$) 
of the population. Separation of the (expected small fraction of)
RQ AGN from the SFGs will be hard for $S_{\rm 1.4GHz} \gtrsim
1~\mu$Jy and impossible at fainter flux densities. At brighter radio flux
densities classification will be possible but not on the very larges areas
covered, e.g., by the EMU survey. Note that the SKA will provide the
sub-arcsec resolution essential for disentangling emission from SF and AGN
activity \citep[e.g.,][]{McAlpine_2015}. Nevertheless, this will not solve the
classification problem if both coexist in RQ AGN, as discussed in
Sect. \ref{sec:emission_mech}.

As regards the optical/NIR bands, the HUDF reaches AB $\sim 29$ ($B$ to
$z$) over 11 arcmin$^2$. The Wide-Field InfraRed Survey Telescope (WFIRST),
a NASA space mission under study for launch in 2024, includes a wide-field
NIR camera to perform surveys with HST-style imaging and sensitivity, but
further in the IR and with hundreds of times the sky coverage
\citep{Gehrels_2015}. The Large Synoptic Survey Telescope
(LSST)\footnote{http://www.lsst.org}, which will be located in Chile, will
provide a survey of about half the sky down to $R_{\rm mag} \sim 27.5$
during 10 years of operation, starting around 2021. This means that the
bulk of the $1~\mu$Jy population (which should have radio-to-optical flux
density ratios $R \lesssim 10$) should be detected by the LSST, i.e., it
will have a counterpart in a large area survey, and will be well within the
sensitivity of WFIRST.  For fainter radio samples optical magnitudes should
get fainter, which will make things more difficult, but the precise values
depend also on the role that dwarf galaxies and low power ellipticals will
play, which is hard to predict (Sect. \ref{sec:future_more}).

Sources having $R_{\rm mag} > 27.5$ will be within reach of the James Webb
Space Telescope (JWST)\footnote{http://www.stsci.edu/jwst/}, due for launch
in 2018, and the Extremely Large Telescopes (ELTs)\footnote{These include,
  in order of decreasing diameter size, the European Extremely Large
  Telescope (E-ELT; http://www.eso.org/sci/facilities/eelt/), the Thirty
  Meter Telescope (TMT; http://www.tmt.org/), and the Giant Magellan
  Telescope (GMT; http://www.gmto.org/).}, with diameters between 25 and 39
m and ``first light'' expected in the mid-2020s, which however will be
covering relatively small fields of view (up to $\approx 0.2$ deg$^2$ for
the smallest ELT). It might turn out that WFIRST, JWST, and the ELTs will
be the main (only?) facilities to secure optical counterparts of nJy radio
sources.

Finally, as regards photometric/spectroscopic redshifts, many of the
$S_{\rm 1.4GHz} \gtrsim 1~\mu$Jy sources might be too faint for current
8/10 m telescopes to be able to provide a redshift and the situation will
get worse at fainter flux densities. This means that JWST and the ELTs
might be the main facilities to secure redshifts of $\mu$Jy radio sources
but even they could have problems in the nJy regime. Redshifts could also
be obtained through radio HI observations. For example, in 10,000 hours 
SKA1 should find
$\approx 5$ million HI galaxies up to $z \sim 0.5$, while SKA2
should detect $\approx 1$ billion HI galaxies up to $z \sim 2$
\citep{Abdalla_2015,Santos_2015}.

I note that \cite{Prandoni_2015} have listed facilities and surveys, which
will complement the proposed SKA1 surveys. These include imaging and
spectroscopic surveys from optical to FIR wavelengths (see their Table 5).

\subsection{The astrophysical impact of future radio observations}

The new radio facilities described in Sect. \ref{sec:new_facilities} will
undoubtedly revolutionise radio astronomy and will have a huge impact on
astrophysics. Readers interested in the research the SKA will foster, for
example, can consult the $\sim 2,000$ page volume Advancing Astrophysics
with the Square Kilometre Array\footnote{Available at
  https://www.skatelescope.org/books/.}. Here I want to give a small 
(somewhat biased) flavour of the topics where we can expect major advances
in the next few years.

\subsubsection{SFGs and cosmic star formation history}\label{sec:future_SFG}

In Sect. \ref{sec:astrop_SFG} I have discussed the importance of studying
the cosmic SF history and how, so far, radio surveys have only played a
marginal role in this field, for reasons to do mostly with
sensitivity. This is going to change quite soon, as detailed in
\cite{Jarvis _2015}.  A 100 nJy limit at 1.4 GHz, for example, would detect
a galaxy with a SFR $\sim 20~\rm M_{\odot}$ yr$^{-1}$ at $z \sim 7$,
pushing the radio band at the forefront in terms of sensitivity to SFR in
comparison to other bands \citep[see Fig. 1a of][]{Madau_2014}. Said
differently, the SKA might provide the most robust measurement of the SF
history of the Universe. And a $\sim 1~\mu$Jy limit, easily reachable by
some of the SKA precursors, would imply a sensitivity to $\sim 50 - 100~\rm
M_{\odot}$ yr$^{-1}$ at $z \sim 6$ \citep{Jarvis _2015}, which is already
very good.

\cite{Madau_2014} quote the difficulty of distinguishing SFGs from AGN in
faint radio surveys as a problem in utilizing radio data but this issue is
under control, at least for $S_{\rm 1.4GHz} \gtrsim 1~\mu$Jy
(Sects. \ref{sec:class} and \ref{sec:class_SKA}). Furthermore, already down
to $S_{\rm 1.4GHz} \sim 30~\mu$Jy the fraction of RL AGN is only $\sim
10\%$ (Fig. \ref{fig:fraction_ECDFS}) and radio emission in sub-mJy RQ AGN
appears to be mainly related to SF processes (Sect.
\ref{sec:emission_mech}). Deeper radio surveys, therefore, are expected to
include mostly SF-related emitters ($\gtrsim 90\%$ for $S_{\rm 1.4GHz}
\gtrsim 1~\mu$Jy: Sect. \ref{sec:predictions}), which would also make the
identification process much simpler and provide a clean(ish) sample for
dealing with this topic.

\subsubsection{Galaxy evolution}\label{sec:future_gal_evol}

A huge amount of effort has been devoted in the past few years to study
galaxy evolution \citep[e.g.,][for a theoretical perspective but also many
  references to observational work]{Somerville_2015}. Radio astronomy
should play a strong\-er role in it \citep[although it has provided jet-mode
  feedback to the modellers: e.g.,][]{Croton_2006}. As mentioned in
Sect. \ref{sec:astrop_RL}, \cite{Best_2014} has been one of the very few
papers to link the observed evolution of jet-mode RL AGN to that of
quiescent galaxies. Future radio surveys will provide us with plentiful
data and it is imperative that a stronger connection is built between the
evolution of radio sources and that of the general population of galaxies,
as has been done, for example, for the RQ AGN population and SFGs
\citep[e.g.][]{Hickox_2014,Caplar_2015}.

We also need to be careful about how to interpret radio data in this
respect. In Sect. \ref{sec:evol_RL} I have discussed the finding by
\cite{Rigby_2015} that the number density of more luminous RL AGN peaks at
redshifts higher than those of lower luminosity objects, as found in many other
bands.  \cite{Yuan_2016} have shown that this (apparently) complex
behaviour displayed by the steep-spectrum radio sources studied by
\cite{Rigby_2015} can be easily reproduced by a simple combination of DE
and LE. The main idea, very simple in retrospect, is that for a population
having a two power-law LF (which is always flat at low powers and steep at
high powers), as is the case for RL AGN, the inferred turnover redshift for
low-luminosity sources will be lower than that of high-luminosity sources,
mimicking a luminosity-dependent density evolution (see their Fig. 3). The very strong
implication is that there appears to be no need for different evolution for
the low- and high-power RL AGN, which has been the mantra in radio
astronomy for many years. This work deserves to be
followed up.

\subsubsection{Why do RL AGN exist?}\label{sec:future_rl_rq}

The first quasars to be discovered were very strong radio sources. More
than fifty years later, we have realised that most AGN are not. But we still do
not know why! Said differently, the question ``Why do only a minority of
galaxies that contain an AGN have jets?'' is still unanswered. Without
clearing this out first, it is going to be very hard to make progress on
galaxy evolution as a whole from a radio perspective
(Sect. \ref{sec:future_gal_evol}).

Many papers have been devoted to the issue of the possible bimodality of
the radio-to-optical flux density ratio $R$ and/or radio power in quasars
\citep[see][for a recent analysis of, and many references on, this
  topic]{Balokovic_2012}. I do not think that this is the real question to
ask \citep[see also][]{Condon_2013}, and even if a bimodality were to be
found it would not answer the fundamental question ``Are there really two
quasar populations?''\footnote{By saying ``quasar'' I refer here only to
  radiative-mode AGN}. In any case, the answer here is already a definite
``yes'' (see Sect. \ref{sec:RQAGN}).

\begin{figure}
\centering
\includegraphics[width=8.4cm]{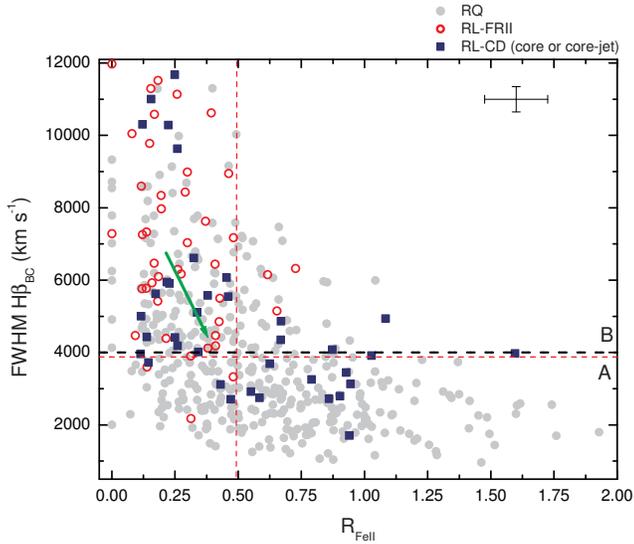}
\caption{RL and RQ quasars in the optical plane of the 4DE1 parameter
  space. RL-FR IIs are double-lobed quasars, i.e., SSRQs, while RL-CD are
  core-dominated quasars, i.e., FSRQs. The green arrow indicates the
  displacement between the median FWHM(H$\beta_{\rm BC}$) and the median
  $R_{\rm Fe II}$ for the two RL classes. The solid light grey symbols are
  RQ objects. In the upper right corner are indicated the typical 2$\sigma$
  errors. The red dotted lines show the boundaries for the RL/RQ separation
  based on a 2D Kolmogorov - Smirnov test. Figure reproduced from
  \cite{Zamfir_2008}, Fig. 4, with permission. }
\label{fig:zamfir}       
\end{figure}

The basic issue, therefore, is not if there are or not two AGN populations
but why, which takes us back to my original question.
There appears to be some interesting differences between RL and RQ AGN,
which should help us in answering this question \citep[see also][for a
  detailed discussion of the first two topics below for $z < 0.7$
  sources]{Tadhunter_2016}. Namely:

\begin{itemize}

\item {\it Environment/Mergers}. RGs appear to be significantly more
  clustered than normal galaxies. In a sphere of 2 Mpc centred on the RG,
  the galaxy density is 2.7 times greater than around normal galaxies
  \citep[e.g.,][]{Velzen_2012}. And \cite{Chiaberge_2015} find that, at $z
  > 1$, $\sim 92\%$ of RGs are associated with recent or ongoing merger,
  while for matched RQ samples this fraction is only $\sim 38\%$.

\item {\it Host galaxy (type and mass)}. Apart from the differences in host
  galaxies between RL and RQ AGN discussed in Sect. \ref{sec:RQAGN}, the
  fraction of galaxies that host RL AGN with $P_{\rm 1.4GHz} > 10^{23}$ W
  Hz$^{-1}$ is a strong function of $\rm M_{\star}$, rising from $\sim 0$
  for $\rm M_{\star} < 10^{10} ~M_{\odot}$ to $\gtrsim 30\%$ at $\rm
  M_{\star} > 5 \times 10^{11}~M_{\odot}$, with a very strong dependence
  $\propto \rm M_{\star}^{2.5}$ \citep[e.g.,][]{Best_2005b}. Given the
  well-known correlation between $\rm M_{\rm BH}$ and $\rm M_{\star}$
  \citep[or more likely $\rm M_{\star,bulge}$: e.g.,][]{Kormendy_2013},
  this implies also that the black hole masses of RL AGN are larger than
  those of RQ AGN.

\item {\it Optical properties.} In a series of papers, Sulentic, Mar\-zia\-ni
  and collaborators, building also on previous work, have proposed a
  four-dimensional parameter space (4D eigenvector 1 [4DE1] based on
  optical, UV and X-ray spectroscopic properties) in an effort to unify
  quasar diversity and an alternate population A--B dichotomy
  \citep[e.g.,][and references there\-in]{Sulentic_2011}. The optical plane
  of this 4DE1 parameter space involves the FWHM of broad H$\beta$ and the
  ratio of the equivalent widths of the Fe II $\lambda$4570 blend and of
  the broad H$\beta$ line ($R_{\rm Fe II}$). Pop. A (FWHM(H$\beta)_{\rm BC}
  < $ 4,000 km s$^{-1}$) is largely RQ, while Pop. B (FWHM(H$\beta)_{\rm
    BC} > $ 4,000 km s$^{-1}$) includes most RL sources and a significant
  number of spectroscopically indistinguishable RQ objects. This suggests
  that RL quasars show significant structural and kinematic differences
  from the majority of RQ sources. The interpretation of the distribution
  of sources on the optical plane is that the average $L/L_{\rm Edd}$
  increases with $R_{\rm Fe II}$, while the dispersion in
  FWHM(H$\beta)_{\rm BC}$ at fixed $R_{\rm Fe II}$ is largely an
  orientation effect \citep[see also][]{Shen_2014}. In addition, the RL
  population shifts to larger FWHM(H$\beta)_{\rm BC}$ and lower $R_{\rm Fe
    II}$ compared to the RQ population, albeit with a large overlap with
  RQ sources, as shown in Fig. \ref{fig:zamfir}
  \citep[from][]{Zamfir_2008}. This is consistent with the notion that RL
  quasars preferentially reside in more massive and lower $L/L_{\rm Edd}$
  systems.

\end{itemize}

Black hole spin has also been suggested to be different between RL and RQ
AGN \citep[e.g.,][]{Wilson_1995,Garofalo_2010}. Unfortunately, reliable
measurements of the black hole spin in AGN are still not available for
sizeable and well-selected samples.

RL AGN appear then to be more clustered, undergoing mergers, reside in
more massive, bulge-dominated galaxies, display broader H$\beta$, and have
lower $L/L_{\rm Edd}$ (and perhaps spin faster or in any case differently)
than RQ AGN. As usual in these cases it would be important to consider
the various, sometimes subtle, selection effects that can plague these
studies. Nevertheless, if these differences are real, the reason why their
combination might explain the presence of jets is still not clear (at
least to me!).

We {\it have} to take advantage of the flood of radio-selected AGN we are
going to discover in the near future to tackle, once and for all, this problem. 

\subsubsection{And more}\label{sec:future_more}

Based on Sect. \ref{sec:predictions} I think that our understanding of what
types of sources to expect down to $S_{\rm 1.4GHz} \approx 1~\mu$Jy is
relatively robust. Below that flux density our vision is more blurred, for
obvious reasons. There are however some populations, which are noticeably
missing from the simulations but that we know have to be there\footnote{I have
discussed two of them in \cite{Padovani_2011a} and summarise the main
points here.}.

\vspace{\baselineskip}
{\it Known unknowns.}

The first population is that of low radio power ellipticals. It has been
known for quite some time that ellipticals of similar optical luminosity
vary widely in radio power. For example, \cite{Capetti_2009} have shown
that 82\% of early-type galaxies in the Virgo cluster with $B_{\rm T} <
14.4$ are undetected at a flux density limit of $\sim 0.1$ Jy, which
implies core radio powers $P_{\rm 8.4GHz} < 4 \times 10^{18}$ W Hz$^{-1}$.
More recently, \cite{Nyland_2016} have studied with the JVLA the nuclear
radio emission of a representative subset of the ATLAS$^{\rm 3D}$ survey of
local early-type galaxies. They detected $51\%$ of their galaxies down to a
5 GHz limit of $\sim 75~\mu$Jy with radio powers as low as $10^{18}$ W
Hz$^{-1}$. These radio faint ellipticals are {\it not} represented in
models of the sub-$\mu$Jy sky: for example, the lower limit of the RL AGN
LF in \cite{Wilman_2008} corresponds to $P_{\rm 1.4GHz} \sim 3 \times
10^{20}$ W Hz$^{-1}$. Sources with lower powers have $S_{\rm 1.4GHz} >
1~\mu$Jy only for $z < 0.3$ (Fig. \ref{fig:flux_z}); and a $P_{\rm 1.4GHz}
= 10^{18}$ W Hz$^{-1}$ object at $z=1$, for example, will have $S_{\rm
  1.4GHz} \sim 0.2$ nJy.

The other missing population is that of dwarf galaxies, which are very
faint and constitute {\it the most numerous} extragalactic population. This
class includes dwarf spheroidals and ellipticals, dwarf irregulars, and
blue compact dwarf galaxies, which all have radio powers reaching $P_{\rm
  1.4GHz} < 10^{18}$ W Hz$^{-1}$, i.e., below the lower limit of the SFG LF
of the simulations of \cite{Wilman_2008}, which by construction do not
include dwarf/irregular galaxies.

The (scanty) available data imply flux density limits $\approx 1$ nJy for
both low-power ellipticals and dwarfs, with surface densities comparable to
those of RQ AGN and SFGs respectively. Both these classes should therefore
be playing a major role in the sub-$\mu$Jy sky but at present it is hard to
be more specific.

\vspace{\baselineskip}
{\it Unknown unknowns.}

There is also what we do not know, or do not expect to find, or still do
not understand, which could be substantial especially at very faint flux
densities.

The balloon-borne Absolute Radiometer for Cosmology, Astrophysics and
Diffuse Emission \citep[ARCADE2;][] {Fixsen_2011} experiment has measured a
sky brightness temperature at 3 GHz $\sim 5$ times that expected from known
populations of radio sources \citep{deZotti_2010,Seiffert_2011}. This means
that either there is a population of discrete radio sources with properties
somewhat different from those of the faint end of the distribution of known
sources or residual emission from our own Galaxy has not been modelled
properly \citep{Seiffert_2011}. In the former case, this new population has
to be exceptionally numerous ($ > 10^{13}$ over the all sky), not
associated with known galaxies, and with $S_{\rm 1.4GHz} \lesssim 0.03~\mu$Jy
\citep{Condon_2012}. In short, there might be still room for some surprises. 

Finally, as stressed by \cite{Norris_2013}, whenever a new facility has
sampled unexplored parts of the observational phase space (be it in
frequency, resolution, time domain, area of sky, etc.), many unexpected
great discoveries in astronomy have been made, very often not by testing a
hypothesis, but just by observing the sky in an innovative way and with an
open mind. I am sure the sub-$\mu$Jy sky will be no exception.

\section{Conclusions}

At this point, it should be clear to all astronomers that the faint radio
sky plays a vibrant role in a variety of astrophysical topics, including
the cosmic star formation history, galaxy evolution, the existence of
powerful jets, and radio emission in RQ AGN. This role will grow even
further in the near future. Radio observations are also unaffected by
absorption, which means, for example, that they are sensitive to all types
of AGN, irrespective of obscuration and orientation (i.e., Type 1s and Type
2s).

I conclude this review by sending the following:

\subsection{Messages to all astronomers}

\begin{enumerate}

\item Do not assume that {\it radio-detected} means {\it radio-loud}. While
  this was almost always true when radio surveys only reached the $\approx$
  Jy level, this is no longer the case, quite the opposite: a sub-mJy AGN
  is more likely to be RQ than RL!

\item The ``radio-quiet AGN'' label is obsolete, misleading, and wrong. The
  major difference between RL and RQ AGN is, based on the available evidence, the presence or
  lack of {\it strong} (relativistic) jets, which in practice translates
  into them being off or on the FIR -- radio correlation. I therefore
  propose that we start using "jetted AGN" and "non-jetted" AGN. This name
  has been used already (albeit very sparsely) in the
  literature\footnote{At the time of writing (mid-2016) I have found 16
    refereed papers with the words "jetted AGN" in their abstract.}. I
  think it is high time it becomes the norm.

\item Do not look for a bimodality in $R$ or $P_{\rm r}$ in quasars, as we
  already know that there are two main classes of AGN: jetted and
  non-jetted.

\item Most importantly: radio astronomy is not a ``niche'' activity
  but is extremely relevant to a whole range of extragalactic studies
  related, for example, to star formation and galaxy evolution. Take
  advantage of that and use radio data!

\end{enumerate}

\subsection{Messages to radio astronomers}

\begin{enumerate}

\item The flattening of deep normalized radio counts is not an open
  issue: we have sorted out the source population of the sub-mJy GHz radio
  sky and learnt that below $\approx 0.1$ mJy the radio sky is dominated by
  SF related processes. This process took more than thirty years because the
  multi-wavelength data necessary to properly classify sources were not
  available and radio astronomy was ahead of the other bands. History does
  not have to repeat itself, although there is some risk that this might
  happen. This can be avoided if radio astronomers understand that ...

\item ... radio astronomy is now a fully multi-wavelength enterprise. This
  is also evinced by the fact that about one third of the papers referenced
  in this review are {\it not} radio papers. The full, proper exploitation
  of data from the SKA and its precursors will {\it require} (this is not
  an option!) synergy with other contemporaneous astronomical
  facilities. These will include, among others, {\it Athena}, WFIRST, the
  LSST, the ELTs, JWST, and SPICA.
  
\end{enumerate}

\begin{acknowledgements}
I thank Roberto Assef, Angela Bongiorno, Ales\-sandro Capetti, Renato
Falomo, Luigina Feretti, Roberto Gilli, Paolo Giom\-mi, Chris Hales,
Evanthia Hatziminaoglou, Darshan Kakkad, Rob\-ert Laing, Vincenzo Mainieri,
Arne Rau, Gordon Richards, an anonymous referee, and particularly Ken Kellermann, for helpful
comments and discussions, and the rest of the E-CDFS team, especially
Margheri\-ta Bonzini, Neal Miller, and Paolo Tozzi, for the work done
together over the past few years. Isabella Prandoni kindly provided me with
most of the data points in Fig. \ref{fig:counts_obs} and the simulated
number counts from the SKADS. I also greatly benefited from the NASA's
Astrophysics Data System.
\end{acknowledgements}

\let\abbrev\nomenclature   \renewcommand{\nomname}{Abbreviations}
\addcontentsline{toc}{section}{Abbreviations}

\nomenclature{6dFGS}{6 degree Field Galaxy Survey}
\nomenclature{AGN}{Active Galactic Nuclei} 
\nomenclature{ARCADE2}{Absolute Radiometer for Cosmology, Astrophysics and Diffuse Emission}
\nomenclature{ALMA}{Atacama Large Millimeter/submillimeter Array}
\nomenclature{BL Lacs}{BL Lacertae objects}
\nomenclature{CSS}{Compact steep-spectrum}
\nomenclature{E-CDFS}{Extended Chandra Deep Field South}
\nomenclature{E-ELT}{European Extremely Large Telescope}
\nomenclature{ELT}{Extremely Large Telescope}
\nomenclature{EMU}{Evolutionary Map of the Universe}
\nomenclature{ESA}{European Space Agency}
\nomenclature{FIR}{Far-IR}
\nomenclature{FR}{Fanaroff-Riley}
\nomenclature{FSRQ}{Flat-spectrum radio quasar}
\nomenclature{FWHM}{Fullwidth half maximum}
\nomenclature{GMT}{Giant Magellan Telescope}
\nomenclature{GOODS}{Great Observatories Origins Deep Survey}
\nomenclature{GPS}{GHz peaked-spectrum}
\nomenclature{HERG}{High-excitation radio galaxy}
\nomenclature{HUDF}{Hubble Ultra Deep Field}
\nomenclature{IR}{Infrared}
\nomenclature{IRAC}{Infrared Array Camera}
\nomenclature{IRAS}{Infrared Astronomical Satellite}
\nomenclature{ISO}{Infrared Space Observatory}
\nomenclature{JVLA}{Karl G. Jansky Very Large Array}
\nomenclature{JWST}{James Webb Space Telescope}
\nomenclature{Jy}{Jansky}
\nomenclature{LERG}{Low-excitation radio galaxy}
\nomenclature{LF }{Luminosity function}
\nomenclature{LOFAR }{LOw Frequency ARray}
\nomenclature{LSST }{Large Synoptic Survey Telescope}
\nomenclature{MIR}{Mid-IR}
\nomenclature{NIR}{Near-IR}
\nomenclature{NRAO}{National Radio Astronomy Observatory}
\nomenclature{NVSS}{NRAO VLA Sky Survey}
\nomenclature{PACS}{Photoconductor Array Camera and Spectrometer}
\nomenclature{PDE}{Pure density evolution}
\nomenclature{PLE}{Pure luminosity evolution}    
\nomenclature{R}{Radio-to-optical flux density ratio} 
\nomenclature{RG}{Radio Galaxy}
\nomenclature{RL}{Radio-loud}
\nomenclature{RQ}{Radio-quiet}
\nomenclature{SB}{Starburst galaxy}   
\nomenclature{SED}{Spectral energy distribution}    
\nomenclature{SDSS}{Sloan Digital Sky Survey} 
\nomenclature{SF}{Star formation}   
\nomenclature{SFG}{Star-forming galaxy}   
\nomenclature{SFR}{Star formation rate}    
\nomenclature{SFRD}{Star formation rate density}   
\nomenclature{SKA}{Square Kilometre Array}    
\nomenclature{SKADS}{SKA Design Study}   
\nomenclature{S/N}{Signal-to-noise}  
\nomenclature{SPICA}{Space Infrared Telescope for Cosmology and Astrophysics}   
\nomenclature{SSRQ}{Steep-spectrum radio quasar}   
\nomenclature{SUMSS}{Sydney University Molonglo Sky Survey}   
\nomenclature{3CRR}{Third Cambridge Catalogue of Radio Sources}   
\nomenclature{TMT}{Thirty Meter Telescope}  
\nomenclature{UV}{Ultraviolet}
\nomenclature{VLA}{Very Large Array}
\nomenclature{VLBI}{Very long baseline interferometry} 
\nomenclature{WFIRST}{Wide-Field InfraRed Survey Telescope}   

\printnomenclature

\addcontentsline{toc}{section}{References}

\small


\end{document}